\documentclass{article}

\usepackage{arxiv}
\usepackage[utf8]{inputenc}                 
\usepackage[T1]{fontenc}                    
\usepackage{amsmath}                        
\usepackage{hyperref}                       
\usepackage{url}                            
\usepackage{booktabs}                       
\usepackage{amsfonts}                       
\usepackage{nicefrac}                       
\usepackage{microtype}                      
\usepackage{cleveref}                       
\usepackage{authblk}                        
\usepackage{graphicx}
\usepackage{caption}                        
\usepackage{subcaption}                     
\usepackage{float}                          
\usepackage[nolist]{acronym}                
\usepackage{multirow}                       
\usepackage{makecell}                       
\usepackage{siunitx}                        
\usepackage[dvipsnames, table]{xcolor}      
\usepackage[numbers, sort&compress]{natbib} 
\usepackage{enumitem}                       
\usepackage{doi}
\usepackage{subfiles}                       

\newcommand{\std}[1]{\scriptsize \num{\pm #1}}

\newcommand{\beginsupplement}{
    \appendix
    \newpage
    \setcounter{table}{0}
    \renewcommand{\thetable}{S\arabic{table}}
    \renewcommand{\theHtable}{S\arabic{table}} 
    \setcounter{figure}{0}
    \renewcommand{\thefigure}{S\arabic{figure}}
    \renewcommand{\theHfigure}{S\arabic{figure}}
    \setcounter{section}{0}
    \renewcommand{\thesection}{S\arabic{section}}
    \renewcommand{\theHsection}{S\arabic{section}}
    \setcounter{equation}{0}
    \renewcommand{\theequation}{S\arabic{equation}}
    \renewcommand{\theHequation}{S\arabic{equation}}
    \setcounter{footnote}{0}
    \acresetall 
}

\title{Robust Quantum Dots Charge Autotuning using Neural Network Uncertainty}

\newbox{\orcid}\sbox{\orcid}{\includegraphics[scale=0.06]{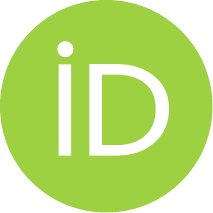}} 

\author[1,2,3]{%
    \href{https://orcid.org/0000-0003-4517-5042}{\usebox{\orcid}\hspace*{1mm}Victor~Yon
    \thanks{\texttt{victor.yon@usherbrooke.ca}}}%
}
\author[1,2,3]{ \href{https://orcid.org/0009-0005-0384-1109}{\usebox{\orcid}\hspace*{1mm}Bastien~Galaup} }
\author[2,3,4]{ \href{https://orcid.org/0009-0003-5789-3807}{\usebox{\orcid}\hspace*{1mm}Claude~Rohrbacher} }
\author[2,3,4]{ Joffrey~Rivard } 
\author[5]{ \href{https://orcid.org/0000-0002-5244-3474}{\usebox{\orcid}\hspace*{1mm}Clément~Godfrin} }
\author[5]{ \href{https://orcid.org/0000-0002-2145-7590}{\usebox{\orcid}\hspace*{1mm}Ruoyu~Li} }
\author[5]{ Stefan~Kubicek } 
\author[5]{ \href{https://orcid.org/0000-0002-1314-9715}{\usebox{\orcid}\hspace*{1mm}Kristiaan~De~Greve} }
\author[6]{ \href{https://orcid.org/0000-0002-1929-2715}{\usebox{\orcid}\hspace*{1mm}Louis~Gaudreau} }
\author[2,3,4]{ Eva~Dupont-Ferrier }
\author[1,2,3]{ \href{https://orcid.org/0000-0003-0311-8840}{\usebox{\orcid}\hspace*{1mm}Yann~Beilliard} }
\author[7,8]{ \href{https://orcid.org/0000-0002-5505-8176}{\usebox{\orcid}\hspace*{1mm}Roger~G.~Melko} }
\author[1,2,3]{ \href{https://orcid.org/0000-0003-2156-967X}{\usebox{\orcid}\hspace*{1mm}Dominique~Drouin} }

\affil[1]{Institut Interdisciplinaire d’Innovation Technologique (3iT), Université de Sherbrooke, Sherbrooke, QC,
    Canada, J1K~0A5}
\affil[2]{Laboratoire Nanotechnologies Nanosystèmes (LN2) — CNRS 3463, Université de Sherbrooke, Sherbrooke, QC, Canada,
    J1K~0A5}
\affil[3]{Institut quantique (IQ), Université de Sherbrooke, Sherbrooke, QC, Canada, J1K~2R1}
\affil[4]{Département de physique, Université de Sherbrooke, Sherbrooke, QC, Canada, J1K~2R1}
\affil[5]{IMEC, Kapeldreef 75, 3001 Leuven, Belgium}
\affil[6]{National Research Council Canada, Quantum and Nanotechnologies Research Center, ON, Ottawa, Canada, K1A~0R6}
\affil[7]{Department of Physics and Astronomy, University of Waterloo, Waterloo, ON, Canada, N2L~3G1}
\affil[8]{Perimeter Institute for Theoretical Physics, Waterloo, ON, Canada, N2L~2Y5}

\renewcommand{\shorttitle}{Robust QD Charge Autotuning using NN Uncertainty}

\hypersetup{
    pdftitle={Robust quantum dots charge autotuning using neural network uncertainty},
    pdfsubject={cs.AI, cs.LG, quant-ph},
    pdfauthor={Victor Yon, Bastien Galaup, Claude Rohrbacher, Clement Godfrin, Ruoyu Li, Stefan Kubicek,
    Kristiaan De Greve, Louis Gaudreau, Eva Dupont-Ferrier, Yann Beilliard, Roger G. Melko, Dominique Drouin},
    pdfkeywords={artificial neural network, Bayesian neural network, quantum dot, charge autotuning,
    uncertainty estimation},
}

\begin{document}
    \maketitle

    \begin{abstract}

    This study presents a machine learning-based procedure to automate the charge tuning of semiconductor spin qubits with minimal human intervention, addressing one of the significant challenges in scaling up quantum dot technologies.
    This method exploits artificial neural networks to identify noisy transition lines in stability diagrams, guiding a robust exploration strategy leveraging neural network uncertainty estimations.
    Tested across three distinct offline experimental datasets representing different single-quantum-dot technologies, this approach achieves a tuning success rate of over \qty{99}{\percent} in optimal cases, where more than \qty{10}{\percent} of the success is directly attributable to uncertainty exploitation.
    The challenging constraints of small training sets containing high diagram-to-diagram variability allowed us to evaluate the capabilities and limits of the proposed procedure.

    \end{abstract}



    \begin{acronym}
        \acro{BCNN}{Bayesian convolutional neural network}
        \acro{CMOS}{complementary metal-oxide-semiconductor}
        \acro{CNN}{convolutional neural network}
        \acro{FF}{feed-forward neural network}
        \acroindefinite{FF}{an}{a}
        \acro{GaAs-QD}{gallium arsenide}
        \acro{ML}{machine learning}
        \acroindefinite{ML}{an}{a}
        \acro{MOS}{metal-oxide-semiconductor}
        \acro{NN}{neural network}
        \acroindefinite{NN}{an}{a}
        \acro{QD}{quantum dot}
        \acro{QPC}{quantum point contact}
        \acro{SEM}{Scanning electron microscope}
        \acro{SET}{single-electron transistor}
        \acro{Si-OG-QD}{silicon overlapping gate}
        \acro{Si-SG-QD}{silicon split gate}
    \end{acronym}

    \newcommand{\michel}{\acs{Si-SG-QD}}
    \newcommand{\louis}{\acs{GaAs-QD}}
    \newcommand{\eva}{\acs{Si-OG-QD}}


    \section{Introduction}\label{sec:introduction}

Multiple quantum computing technologies compete to outperform classical computers for practical applications, yet the journey toward achieving this objective is fraught with considerable technical challenges.
This article specifically studies quantum systems based on semiconductor spin qubits~\cite{Loss_1998, Veldhorst_2015, Watson_2018, Burkard_2023} formed by electrostatically confining electrons in \acp{QD}.
This technology presents several advantages, such as high gate fidelities~\cite{Takeda_2016, Yoneda_2017, Mills_2022, Noiri_2022, Xue_2022}, long coherence times~\cite{Tyryshkin_2011, Veldhorst_2014}, thermal robustness at temperatures greater than one kelvin~\cite{Petit_2022}, and compatibility with well-established \ac{CMOS} technologies~\cite{Maurand_2016, Stuyck_2021, Elsayed_2024}.
This compatibility further amplifies scalability potential and enables co-integration with industrial electronics~\cite{Gonzalez_Zalba_2021, Rohrbacher_2023}.
However, \ac{QD} technology is still in an early stage of development, and the best materials~\cite{Liu_2019, Saraiva_2021}, fabrication techniques, use cases, and control procedures are yet to be discovered.

One bottleneck in the large-scale implementation of quantum computers based on \acp{QD} is the complexity of charge tuning: one of the preliminary tasks required to set up a qubit.
This tuning task confines a specific number of charge carriers in one or multiple \acp{QD} using the information provided by indirect measurements, which is typically obtained by tuning the voltages at two control gates (G1 and G2 in Figure~\ref{fig:dot-devices}).
The results of voltage sweeping can be visualized as two-dimensional images termed stability diagrams (examples in Figure~\ref{fig:datasets}a,b,c), where pixels encode the current measured using \iac{SET}~\cite{Patel_2020} or \iac{QPC}~\cite{Simmons_2007}.
One could then identify the transition lines (highlighted with green lines in Figure~\ref{fig:datasets}d,e,f) that represent a change in the number of charge carriers inside the \ac{QD}.
Given the knowledge that electron-based \acp{QD} are empty at low gate voltages, it is possible to deduce the number of charges for each area in the voltage space (blue areas in Figure~\ref{fig:datasets}d,e,f).

However, the task is challenging to automate because the stability diagram parameter space is vast, noisy, and device-dependent.
For this reason, the number of charges inside the \ac{QD} is usually tuned manually by experimentalists based on informed guesses and human heuristics.
While this approach is sufficient for small proof-of-concept studies in research laboratories, it is incompatible with large-scale industrial applications.
Moreover, measuring a large area of a stability diagram could take hours, which significantly slows down \ac{QD} initialization, especially since it often requires iterative measurements.
One could attempt to automate the tuning procedure using classical algorithms and human expertise~\cite{Baart_2016}, but this approach still requires human involvement and is tailored to specific hardware.
Thus, it does not satisfy the requirements for scaling up the technology or accelerating the development of new designs.

Recent studies~\cite{Czischek_2021, Durrer_2020} have suggested automating the tuning procedure using \iac{ML} approach with artificial deep \acp{NN}~\cite{LeCun_2015}.
These methods have shown promising results but are not yet reliable enough to consider fully autonomous large-scale \ac{QD} control.
The relatively high failure rate of these \ac{ML} approaches (\qty{25}{\percent}~\cite{Czischek_2021} for single-\ac{QD} and \qty{43}{\percent}~\cite{Durrer_2020} for double-\ac{QD} charge tuning) could be attributed to \emph{(i)} the overconfidence of the model while facing ambiguous, out-of-distribution, too small, or noisy measurements and \emph{(ii)} the low error tolerance of the exploration strategy.
For example, one transition line misdetection will often lead to the failure of the entire charge tuning procedure.
In the context of coarse tuning, \citet{Ziegler_2022} addressed a similar issue by first training a \ac{NN} to evaluate the measurement quality before sending it to a second \ac{NN} to infer the class.
\citet{Ziegler_2023} combined this method with custom peak-finding techniques to tune offline double \acp{QD} into a specific charge regime with an accuracy of \qty{89.7}{\percent}.
Here, we chose a more versatile approach by training a single \ac{NN} to detect the transition lines and providing a classification confidence score at the same time.
One could also improve the performance of \ac{ML} models by using more parameters and a larger dataset, but a very large model could be challenging to integrate near the tuned devices, and the high cost of collecting more diagrams is prohibitive.
Even so, \ac{ML} models never provide the guarantee of perfect accuracy.

This study introduces a hardware-agnostic autotuning procedure that leverages \ac{NN} uncertainty to significantly enhance the robustness of semiconductor spin qubit charge tuning with minimal human intervention.
This is achieved by training \iac{NN} to identify transition lines in a stability diagram using supervised learning, coupled with an exploration strategy incorporating the \ac{NN}'s predictions and confidence score (measure of uncertainty).
We considered two methods to estimate the confidence score of the \acp{NN}: one with a heuristic from classical \acp{NN} and one using the Bayesian framework applied to \acp{NN}~\cite{Goan_2020, Gal_2016}.
These two methods have been evaluated based on their concrete ability to reduce the critical failure rate during the autotuning procedure across three different experimental offline datasets measured using different single-\ac{QD} hardware.

    \section{Problem definition}\label{sec:problem}

    \subsection{Project scope}\label{subsec:scope}

Before using \acp{QD} as qubits to perform operations, each \ac{QD} must be tuned to a specific charge state by applying appropriate gate voltages.
The complete calibration procedure can be divided into five distinct steps: \emph{(i)} \emph{bootstrapping}: cooling the device and bringing its regime into the appropriate parameter range; \emph{(ii)} \emph{coarse tuning}~\cite{Ziegler_2023, Liu_2022, Zwolak_2020, Severin_2024, Kalantre_2019, Darulova_2021, Moon_2020, Ziegler_2022}: tuning the \acp{QD} to a specific topology (e.g., single \ac{QD}, double \ac{QD}); \emph{(iii)} \emph{establishing controllability}~\cite{Ziegler_2023, Liu_2022, Perron_2015, Hensgens_2018}: setting up virtual gates that compensate for capacitive cross-talk; \emph{(iv)} \emph{charge state tuning}~\cite{Ziegler_2023, Czischek_2021, Lapointe_Major_2020, Durrer_2020}: tuning the \acp{QD} to a specific charge configuration (number of electrons in our case); \emph{(v)} \emph{fine-tuning}~\cite{Liu_2022, Teske_2019}: adjusting the inter-dot tunnel coupling.
For a more detailed description of these steps, refer to~\citet{Zwolak_2023}.
Performing these tasks quickly and reliably is one of the technical challenges preventing the scale-up of the number of qubits in \ac{QD}-based quantum computers.

This work focuses on automating the task of charge state tuning \emph{(iv)}.
We assume the system is bootstrapped \emph{(i)} and the voltage range contains a single \ac{QD} \emph{(ii)}.
However, our method does not require setting up virtual gates \emph{(iii)}, since we opted for a flexible approach by detecting the slope of the transition lines during the procedure.

    \subsection{Problem frame and constraints}\label{subsec:frame}

We framed the problem of charge state tuning as an exploration task, where each step consists of detecting charge transition lines in the measurement of a voltage space subsection.
The ultimate goal is to reach a specific charge regime while minimizing the exploration cost.
In this case, the cost is the measurement time, which is directly affected by the number of steps and the size of the subsection measured.
This approach allows us to break down the problem into two distinct tasks: transition line detection (Section~\ref{sec:line}) and the diagram exploration strategy (Section~\ref{sec:autotuning}).

To uphold the robustness and adaptability of our methodology, we consciously minimized the utilization of preprocessing techniques while maintaining a relatively simple \ac{NN} model.
This decision was grounded in the ultimate objective of implementing this solution in custom electronics inside the cryogenic environment of a dilution refrigerator to perform in situ online tuning.
The computational constraints and the need for a streamlined process inherent to this environment necessitate a careful balance between model complexity and operational energy efficiency.
To evaluate the efficacy of our approach across various \ac{QD} designs, we tested our method on three distinct datasets, each obtained from a different hardware and research group.
This ensures that our solution is not tailored specifically to one quantum device but can be applied to a wide range of \ac{QD}-based spin qubit hardware.

    \subsection{Datasets}\label{subsec:datasets}

    \begin{figure}
        \centering
        \begin{subfigure}{0.32\textwidth}
            \centering
            \includegraphics[width=\textwidth]{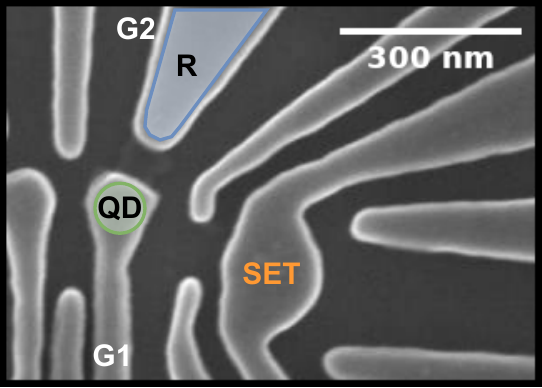}
            \caption{Device \acl{Si-SG-QD}\\(\acs{Si-SG-QD})}
            \label{fig:dot-device-michel}
        \end{subfigure}
        \begin{subfigure}{0.32\textwidth}
            \centering
            \includegraphics[width=\textwidth]{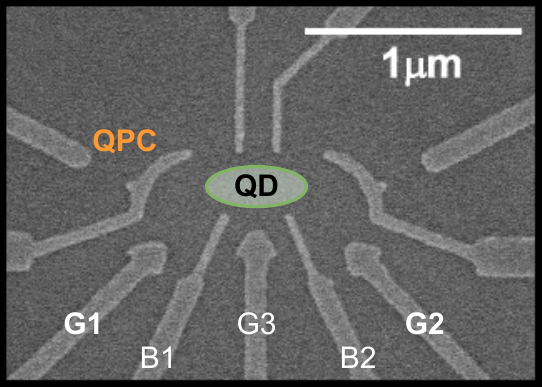}
            \caption{Device \acl{GaAs-QD}\\(\acs{GaAs-QD})}
            \label{fig:dot-device-louis}
        \end{subfigure}
        \begin{subfigure}{0.32\textwidth}
            \centering
            \includegraphics[width=\textwidth]{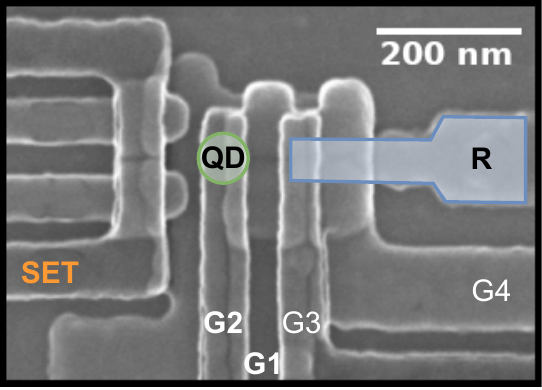}
            \caption{Device \acl{Si-OG-QD}\\(\acs{Si-OG-QD})}
            \label{fig:dot-device-eva}
        \end{subfigure}
        \begin{subfigure}{0.32\textwidth}
            \centering
            \includegraphics[width=\textwidth]{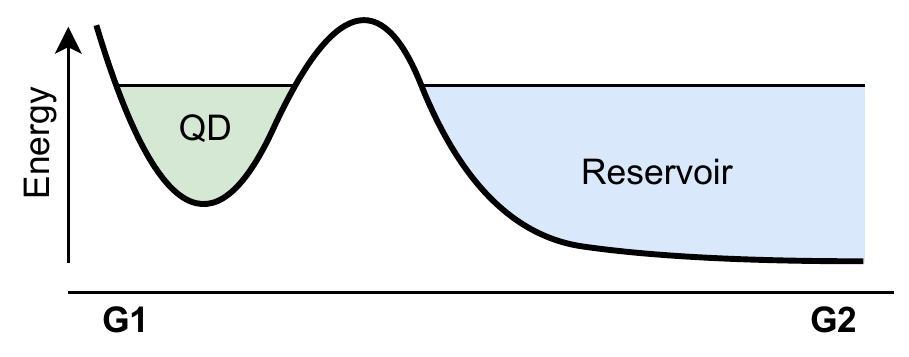}
            \caption{\acs{Si-SG-QD}\\energy diagram}
            \label{fig:dot-energy-michel}
        \end{subfigure}
        \begin{subfigure}{0.32\textwidth}
            \centering
            \includegraphics[width=\textwidth]{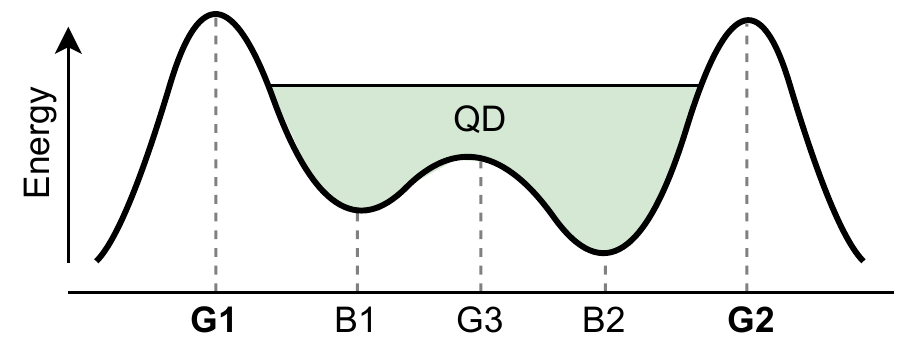}
            \caption{\acs{GaAs-QD}\\energy diagram}
            \label{fig:dot-energy-louis}
        \end{subfigure}
        \begin{subfigure}{0.32\textwidth}
            \centering
            \includegraphics[width=\textwidth]{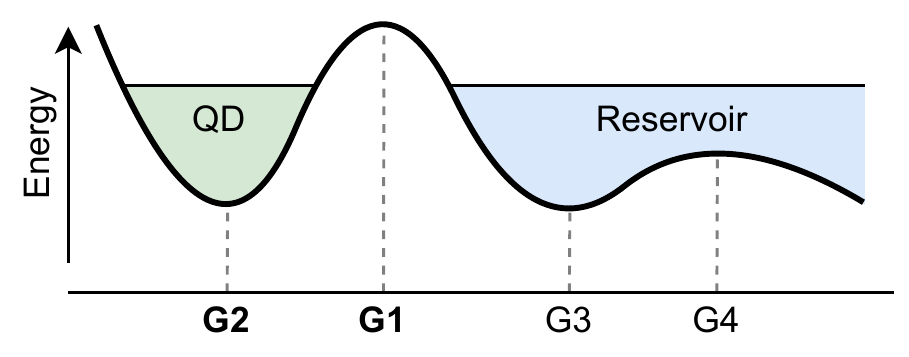}
            \caption{\acs{Si-OG-QD}\\energy diagram}
            \label{fig:dot-energy-eva}
        \end{subfigure}
        \caption[Quantum dot devices and energy diagrams]{
            \Acf*{QD} devices used to measure the stability diagrams. \textbf{(a,b,c)} \acf*{SEM} images of experimental devices.
            A single \ac{QD} can be created by tuning electrostatic gates 1 (G1) and 2 (G2) to appropriate voltages.
            \textbf{(a)} G2 controls the electronic density of an electron reservoir (R), and the \acf*{SET} is used as a charge sensor.
            See~\cite{Rochette_2019} for experimental details.
            \textbf{(b)} The tunnel barriers labeled B1 and B2 are set to a fixed voltage, as well as gate 3 (G3), which is required for more than one \ac{QD} only.
            The \acf*{QPC} is used as a charge sensor.
            See~\cite{Gaudreau_2009} for experimental details.
            \textbf{(c)} G1 and G4 are tunnel barriers.
            The voltages at G3, G4, and R are fixed to allow for single-\ac{QD} formation under G2 and an extended reservoir.
            The \ac{SET} is used as a charge sensor.
            See~\cite{Elsayed_2024} for experimental details.
            \textbf{(d,e,f)} Energy diagrams representing the formation of a single \ac{QD} for each device.
        }
        \label{fig:dot-devices}
    \end{figure}

\begin{figure}
        \centering
        \includegraphics[width=.95\textwidth]{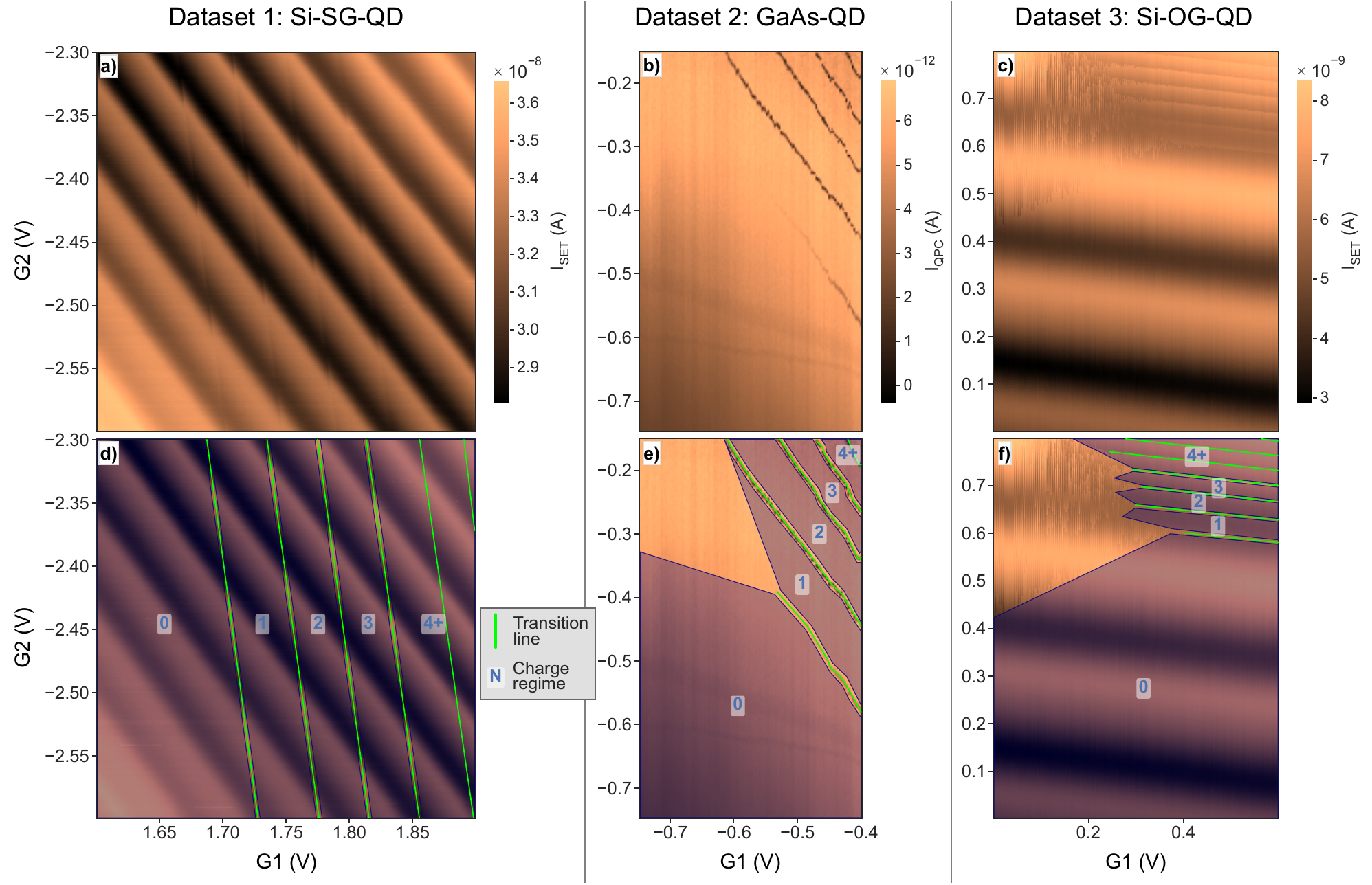}
        \caption[Stability diagram examples]{
            Example of one stability diagram from each dataset.
            \textbf{(a,b,c)} Representation of the stability diagrams as images, where pixels encode the measured current for given voltage values at the gates (G1 and G2).
            \textbf{(d,e,f)} Same diagrams with manual annotations of transition lines in green and charge regime areas in blue.
            The regions with more than \num{3} charges are grouped under the annotation ``\qty{4}{+}''.
            A voltage area not annotated with a charge regime due to fading lines or ambiguous boundaries is considered an ``unknown charge regime''.
            More examples can be found in Supplementary Section~\ref{subsec:digram-samples}.
        }
        \label{fig:datasets}
    \end{figure}

We tested our approach on offline experimental measurements obtained from three different \ac{QD} device architectures across distinct experimental setups.
The first dataset, referred to as \acf{Si-SG-QD}~\cite{Rochette_2019}, contains \num{17} stability diagrams (Figure~\ref{fig:datasets}a) measured from devices fabricated with \iac{MOS} single-layer architecture (Figure~\ref{fig:dot-devices}a).
The second dataset, referred to as \acf{GaAs-QD}~\cite{Gaudreau_2009}, is composed of \num{9} stability diagrams (Figure~\ref{fig:datasets}b) obtained by measuring the current of a device fabricated with a GaAs/AlGaAs heterostructure (Figure~\ref{fig:dot-devices}b).
The third and last dataset, referred to as \acf{Si-OG-QD}~\cite{Elsayed_2024}, includes \num{12} stability diagrams (Figure~\ref{fig:datasets}c) measured from devices fabricated with \iac{MOS} stacked-layer architecture (Figure~\ref{fig:dot-devices}c).
The diagrams for the first and third datasets are measured using \iac{SET}, while the ones for the second dataset are measured using a \ac{QPC}.
The measurements are two-dimensional stability diagrams represented as images, where the $x$- and $y$-axes correspond to the voltages applied to gates G1 and G2, respectively.
If more than two gates are available in the experimental setup, the other ones are fixed to a voltage compatible with a single-\ac{QD} regime.

Using simulated stability diagrams could increase the size of the training set~\cite{Darulova_2021, Zwolak_2018}, but at the cost of a distribution shift between the training and testing sets, which could harm the quality of uncertainty quantification.
We therefore chose to rely exclusively on experimental data.
We also minimized measurement preprocessing to ensure compatibility with future in situ online implementation of this autotuning method.
Only input normalization was applied to the measured patch before being sent to the classification model.
See Supplementary Section~\ref{sec:suppl-datasets} for detailed information regarding dataset preparation and diagram selection.

Each dataset features different defects, noise, and transition line patterns, creating a large variety of challenges for the autotuning procedure.
The diagrams from \ac{Si-SG-QD} present strong oscillating background noise caused by the cross-capacitance effects of G1 and G2 acting on the \ac{SET}, which can be hard to differentiate from the transition lines (see diagonal parasitic oscillations in Figure~\ref{fig:datasets}a).
Therefore, the most challenging part of tuning such devices is reliably detecting the lines, especially in low- and high-current areas.
The challenge is entirely different for diagrams from the \ac{GaAs-QD} dataset, where the transition lines are usually easy to identify but can fade (see first line in Figure~\ref{fig:datasets}b) and curve unexpectedly.
This specificity moves the challenge to the exploration task, where missing a fading line can lead to a wrong number of charges in the \ac{QD}.
Finally, the diagrams from \ac{Si-OG-QD} present a peculiar hysteresis noise when the reservoir tunneling rate is slower than the voltage sweeping speed (see top left in Figure~\ref{fig:datasets}c), which must be correctly classified by the model and taken into account by the exploration strategy.
Supplementary Section~\ref{subsec:digram-samples} presents a diagram sample of each dataset.
All measurement files used to generate the diagrams in the present study can be downloaded from \citet{qdsd}.

    \section{Line detection}\label{sec:line}

    \subsection{Methods}\label{subsec:line-method}

    \begin{figure}
        \centering
        \includegraphics[width=.95\textwidth]{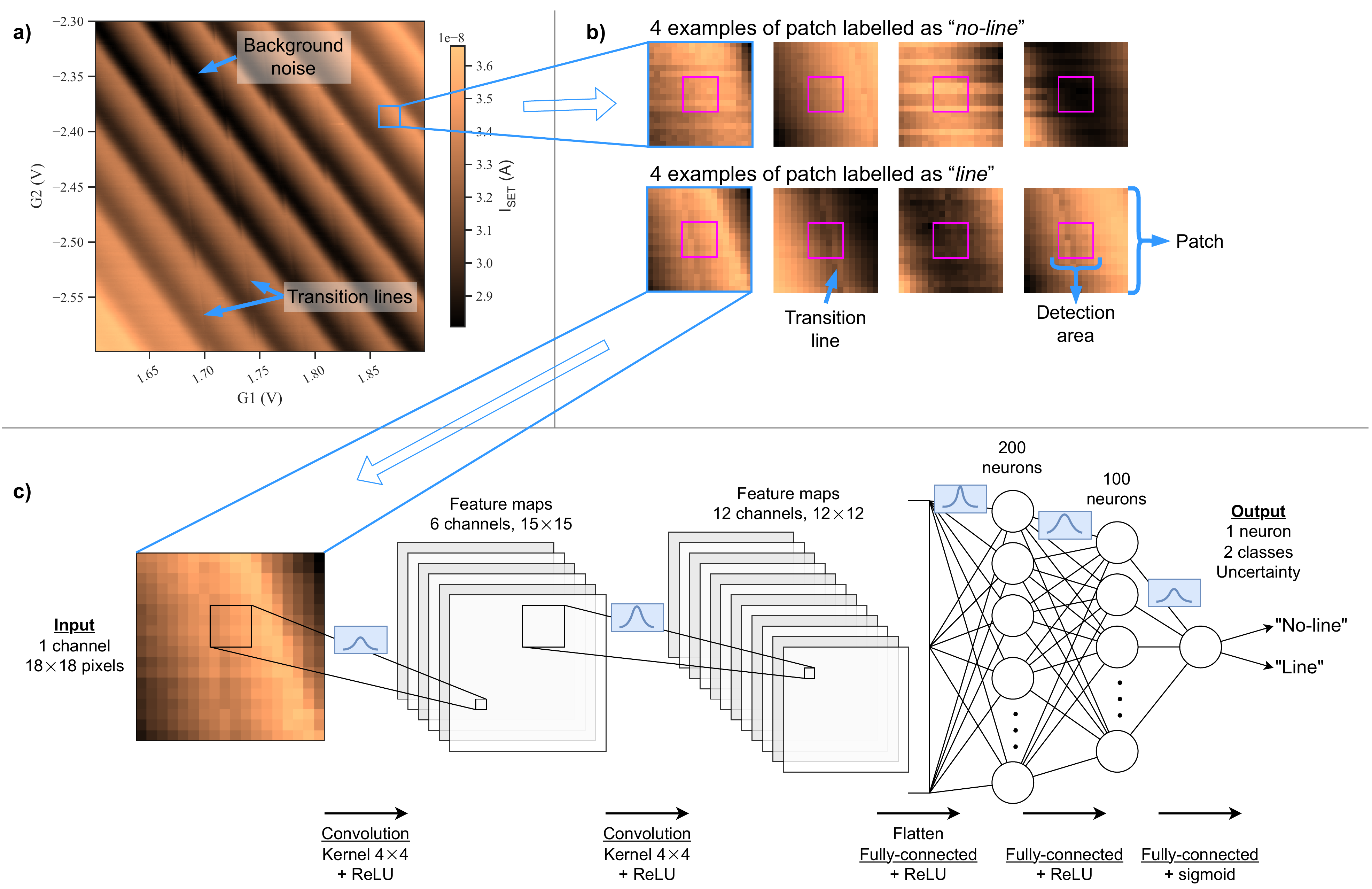}
        \caption[Transition line detection using a neural network]{
            Automatic transition line detection using a \acf*{BCNN} in a stability diagram.
            \textbf{(a)} A subsection (patch) of the voltage space is measured.
            This diagram presents strong oscillating background noise that should not be misinterpreted as a transition line.
            \textbf{(b)} To train the model, each patch is categorized as ``\emph{line}'' if a transition line annotation intersects with the detection area (in pink) in the center of the patch.
            Otherwise, the patch is categorized as ``\emph{no-line}''.
            \textbf{(c)} The patch is then sent as input to a model, where a forward pass propagates the information through convolutional and fully connected layers.
            In this example, a \acs*{BCNN} is represented, where each parameter is encoded as a Gaussian distribution.
        }
        \label{fig:network}
    \end{figure}

The first part of the procedure consists of automatically detecting charge transition lines in subsections of offline experimental stability diagrams (Figure~\ref{fig:network}).
An image segmentation model~\cite{Minaee_2021, Zhang_2024} could identify lines, but it would require scanning large diagram sections, which is expected to slow down the autotuning process.
To relax the input size requirement and reduce the model's complexity, we address this problem as a supervised binary classification task (``\emph{line}'' or ``\emph{no-line}'').
The choice of a \ac{NN} rather than alternative statistical methods~\cite{Ding_2001, Bishop_2006, Mukhopadhyay_2015, Sun_2022} is firstly driven by the goal of designing a hardware-agnostic method that can adapt to the rapidly evolving \ac{QD} research field.
\Ac{ML} algorithms can automatically adjust to new data features, such as a new artifact caused by a novel measurement setup or new transition line patterns induced by a change in the \ac{QD} hardware design.
\Ac{NN} methods also have the potential to employ transfer learning methods~\cite{Bengio_2012} to facilitate swift adaptation of the model to newly introduced devices.
Furthermore, \acp{NN} have demonstrated outstanding performance in detecting patterns within images and filtering various types of noise, especially \acp{CNN}~\cite{Krizhevsky_2017, Patil_2020, Chen_2021, Li_2022}.
Therefore, they are well suited for detecting lines within images, such as transition lines in two-dimensional stability diagrams.
Finally, \acp{NN} are known to maintain good scalability~\cite{Lienhard_2022} with respect to problem size and complexity; a crucial characteristic as we aim to develop a solution that is compatible with large arrays of \acp{QD}.

For this classification problem, we evaluated the performance of three \ac{NN} architectures: \acp{CNN}, \acp{BCNN}, and \acp{FF}.
\acp{CNN} are known for their effectiveness in image classification tasks~\cite{Chen_2021}, and they provide a good tradeoff between complexity and performance.
\acp{BCNN} are expected to offer robust uncertainty estimations~\cite{Kwon_2020, Jospin_2022}, which can be beneficial for improving the reliability of automatic charge-tuning procedures.
\acp{BCNN} implement the variational inference~\cite{Hinton_1993, Graves_2011} method with the Bayes-by-Backprop~\cite{Blundell_2015} learning rule.
Meanwhile, the simpler \ac{FF} is a reference for comparison with the more complex \ac{CNN} and \ac{BCNN} models.

All \acp{NN} presented in this article are fed with the same inputs and have comparable binary outputs.
The model input is a small subsection of the voltage space (also referred to as a patch), where pixels represent the measured current values.
The patch size is fixed to \numproduct{18 x 18} data points as an empirical tradeoff between measurement speed and model accuracy.
The patches are generated by splitting the diagrams into evenly spaced squares and then distributing them into training, validation, and test sets, where the test patches are extracted from a unique diagram excluded from the training set, and the other patches are randomly distributed between the training (\qty{90}{\percent}) and validation (\qty{10}{\percent}) sets.
Each patch is automatically classified as ``\emph{line}'' if an annotation of the transition line intersects with the detection area in its center (Figure~\ref{fig:network}b).
This labeling approach provides more context to the \ac{NN} while keeping the classification window narrow enough to fit between two transition lines.

The most basic \ac{NN} used here is the \ac{FF}, which propagates the input through two fully connected layers.
The \ac{CNN} extends this model by adding two convolution layers before the fully connected layers.
The \ac{BCNN} architecture is identical to the \ac{CNN}, with each free parameter (weights and biases) encoded as a Gaussian distribution, defined by a mean and a variance (Figure~\ref{fig:network}c).
Each model is trained to infer the patch class as a binary output and express the uncertainty of this classification as a confidence percentage.
More technical details about data processing, manual diagram annotation, automatic patch labeling, and \ac{NN} meta-parameters can be found in Supplementary Sections~\ref{sec:suppl-datasets} and~\ref{sec:suppl-line-method}.

Introducing a confidence score into our model provides an additional layer of information about the model's predictions.
We leverage this information at the exploration level to significantly reduce the impact of misclassifications, which are usually responsible for disrupting the autotuning procedure.
We distinguish between three types of uncertainties~\cite{Gawlikowski_2021, Smith_2018}:

\begin{enumerate}
\item
    \nopagebreak Model uncertainty (also known as epistemic or systemic uncertainty) arises from a lack of knowledge due to imperfect training.
    \nopagebreak In the context of transition line detection, this could be due to non-optimal meta-parameters or a low number of training diagrams.
    \nopagebreak This type of uncertainty can be reduced as the model is exposed to more data, highlighting the importance of a comprehensive training process.
    \nopagebreak In practice, the size of the training set is restrained by the acquisition cost of the experimental stability diagrams.
\item
    \nopagebreak Data uncertainty (also known as aleatoric or statistical uncertainty), on the other hand, is associated with the inherently stochastic nature of the tested data.
    \nopagebreak For the line classification task, this could result from variability in measurement tools, fabrication imperfections, or cross-talk, among other factors.
    \nopagebreak These factors are often out of the experimentalist's control.
\item
    \nopagebreak Finally, distributional uncertainty arises from an inadequate representation of the test set within the training set.
    \nopagebreak In the context of line classification, this can occur when the stability diagram used for testing presents noise or features that are not present in the training set.
    \nopagebreak Not using synthetic data helps us mitigate this type of uncertainty, but diagram-to-diagram variability is a source of distributional uncertainty that can never be entirely avoided.
\end{enumerate}

The computation of the confidence score is handled differently between standard (\ac{CNN} or \ac{FF}) and Bayesian \acp{NN}.
For a standard \ac{NN}, we employ a simple heuristic score (Formula~\ref{eq:conf-heuristic}) to estimate uncertainty based on the distance between the output $y$ of the \ac{NN} and the closest class~\cite{Zaragoza_1998, Mandelbaum_2017}.
This estimation is based on the probabilistic interpretation~\cite{Ramos_2018} of the cross-entropy function~\cite{Mao_2023} used to optimize the model.
In the case of \acp{BCNN}, we use the variational inference~\cite{Graves_2011, Gawlikowski_2021} approach by computing the normalized standard deviation $\sigma$ based on $N$ repeated inferences with newly sampled parameters (Formula~\ref{eq:conf-bayes}).
Both formulas are trivial in the case of binary classification with one output neuron, but they can be generalized to multi-class problems~\cite{Zaragoza_1998}.

    \begin{equation}\label{eq:conf-heuristic}
        \text{Heuristic confidence score} = |0.5 - y| \times 2
    \end{equation}

    \begin{equation}\label{eq:conf-bayes}
        \text{Bayesian confidence score} = 1 - \sigma(y_1, y_2, \dots, y_N) \times 2
    \end{equation}

In addition, we performed a calibration step to optimize the exploration--exploitation tradeoff~\cite{Auer_2002, Simpkins_2008}.
\Acp{NN} are typically calibrated using regularization methods during training or by empirically scaling individual bins of the reliability diagram after training~\cite{Gawlikowski_2021, Guo_2017, Vaicenavicius_2019, Filho_2023} to ensure that the confidence score is a good approximation of the actual probability of correctness.
For example, among all classifications rated with \qty{80}{\percent} confidence, \qty{20}{\percent} of them are expected to be incorrect.
In our case, the validation set does not consistently cover the range of confidence scores due to the low number of classification errors in some cases (e.g., the middle panel of Supplementary Figure~\ref{fig:threshold-calibration}b).
This constraint makes the computation of calibration metrics unreliable, specifically for bin-based approaches.
We worked around this problem by calibrating the confidence thresholds instead of the model confidence scores.
The threshold defines the confidence value under which the classification will be considered too uncertain to be trusted during stability diagram exploration.
After each training, the threshold value is optimized by minimizing the score defined in Formula~\ref{eq:threshold} through a grid search on the validation set.
The goal is to find an optimal tradeoff between the number of errors above the threshold ($\text{Err}$) and the total number of samples under the threshold ($\text{UT}$), where $\tau$ is a meta-parameter that defines the target ratio (fixed to \num{0.2} in this study).
This approach relaxes the requirement regarding the number of calibration samples while maintaining the practical functionality of the confidence score.
More details about the calibration process can be found in Supplementary Section~\ref{subsec:threshold}.

    \begin{equation}\label{eq:threshold}
        \text{Threshold score} = \text{Err} + \text{UT} \times \tau
    \end{equation}

    \subsection{Results}\label{subsec:line-results}

With appropriate meta-parameters (Supplementary Table~\ref{tab:model-parameter}), the line classification task reached more than \qty{90}{\percent} accuracy on every dataset and model combination (Table~\ref{tab:line-results}).
The most significant performance difference between the \ac{NN} architectures is visible for the \ac{Si-SG-QD} dataset, where convolution layers are necessary to filter the oscillating parasitic background.
For the \ac{GaAs-QD} and \ac{Si-OG-QD} datasets, the benefits of the convolution are less evident, since the measurement noise is lower and the transition lines are often clearly visible.
For all the datasets, the classification performance of the Bayesian \ac{CNN} is close to that of their classical counterparts, but the additional complexity of the Bayesian layers slows down the training by a factor of \num{4} on average.
No overfitting was observed during training (Supplementary Figure~\ref{fig:train-progress}), suggesting that the dropout of the classical layers and the Bayesian layers both act as efficient regularization methods.

\begin{figure}
        \centering
        \includegraphics[width=.95\textwidth]{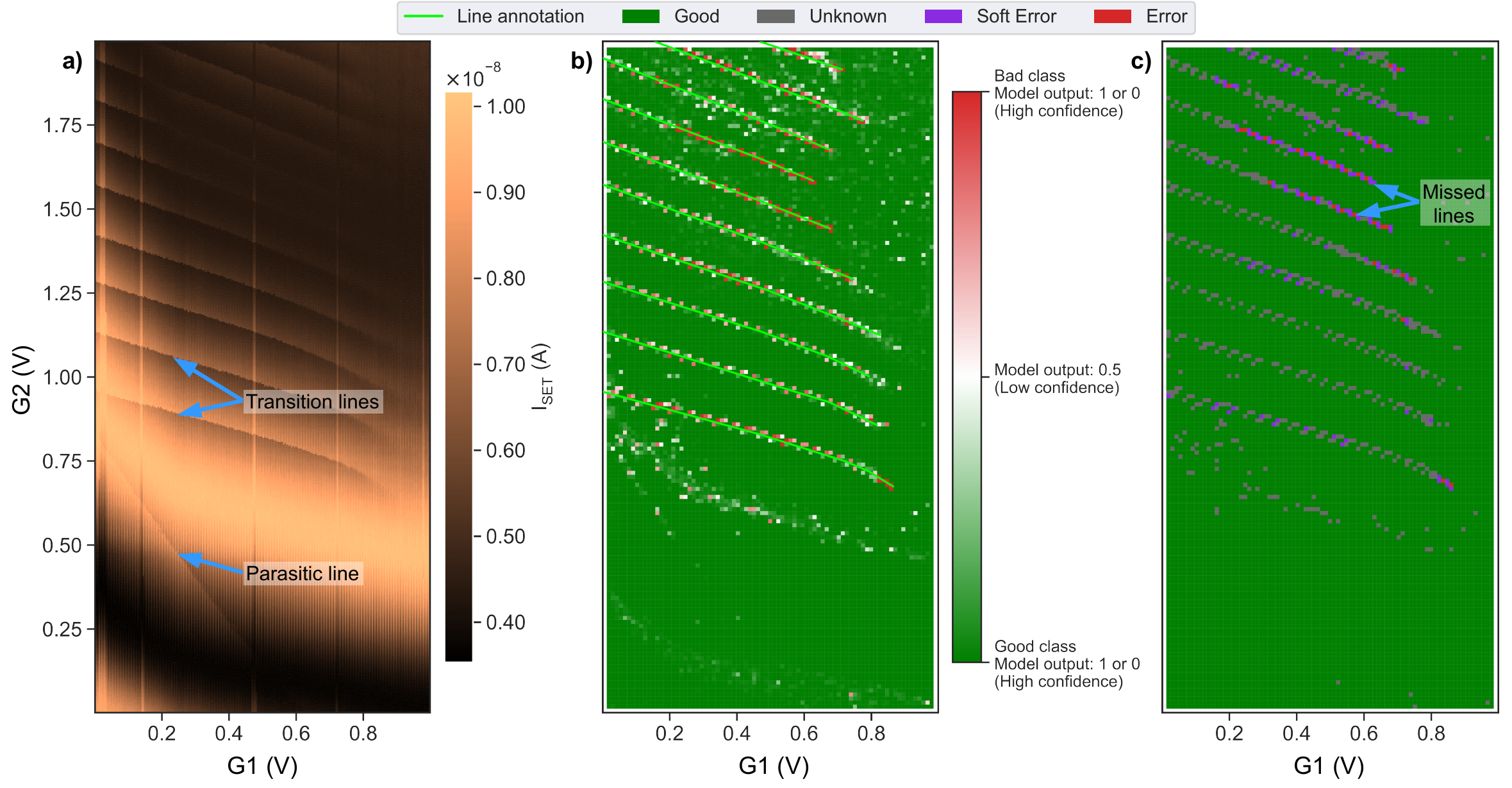}
        \caption[Example of transition line classification for a full stability diagram]{
            \textbf{(a)} Example of a stability diagram from the \ac{Si-OG-QD} dataset.
            This specific diagram presents a parasitic line in the low-voltage area.
            \textbf{(b)} The same stability diagram is divided into patches classified using \iac{CNN} trained using cross-validation~\cite{Raschka_2018} (Supplementary Figure~\ref{fig:cross-validation}).
            The color gradient of the patches represents the model uncertainty, where \num{0.5} is the lowest confidence score according to Formula~\ref{eq:conf-heuristic}.
            The confidence score is lower in low-contrast areas, around the parasitic line, and near transition lines (due to possible intersection ambiguity between the line labels and the detection area of the patch).
            \textbf{(c)} The same stability diagram after the application of the confidence threshold.
            Most errors are avoided, except for the end of a few fading lines and some line sections in the low-current area at the top of the diagram.
            The ``soft errors'' represent misclassifications near a line, which are not expected to induce tuning errors.
        }
        \label{fig:full-map}
    \end{figure}

A qualitative analysis of the misclassified patch samples (Figure~\ref{fig:full-map} and Supplementary Section~\ref{subsec:patch-samples}) suggests that we approach an optimal classification rate for each dataset, even if the model accuracy never reaches \qty{100}{\percent}.
The remaining errors are often caused by ambiguous inputs that would be hard to classify at the patch scale, even by a human expert.
The most common causes of misclassification are as follows: \emph{(i)} ambiguous labeling (a fading line or a line annotation near the detection area), \emph{(ii)} annotation errors (position inaccuracy or human error), and \emph{(iii)} strong noise at the patch location.
The proportion of ambiguous patches varies between the datasets, which explains why the average line classification accuracy differs.
The transition lines from \ac{GaAs-QD} and \ac{Si-OG-QD} can be irregular and fading (\qty{94.4}{\percent} and \qty{92.5}{\percent} accuracy with the best models, respectively), while the \ac{Si-SG-QD} lines are straight and more predictable, simplifying patch classification (\qty{96.9}{\percent} accuracy with the \ac{CNN}).
The most challenging errors to address are the ones related to out-of-distribution issues caused by features in the test diagram that were missing from the training set.
In an online scenario, this issue could occur if the device we are trying to tune is too different or if there are physical defects that did not exist in the previous experiments.
This problem could be mitigated by a more diversified training set, a higher \ac{QD} fabrication quality with less variability, or a reliable confidence score that correctly expresses the distributional uncertainty.

\begin{figure}[H]
            \centering
            \begin{subfigure}{0.4\textwidth}
                \centering
                \includegraphics[width=\textwidth]{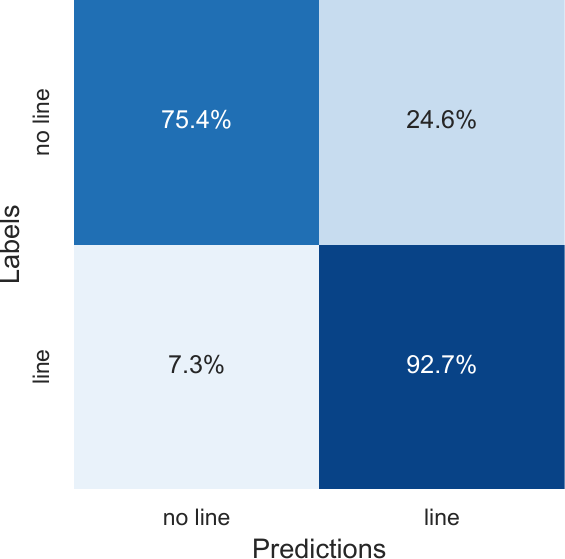}
                \caption{
                    Confusion matrix without using the confidence threshold.
                    Accuracy: \qty{81.5}{\percent}; precision: \qty{81.0}{\percent}; recall: \qty{84.1}{\percent}; F1 score: \qty{80.9}{\percent}.
                }
            \end{subfigure}\hfill
            \begin{subfigure}{0.56\textwidth}
                \centering
                \includegraphics[width=\textwidth]{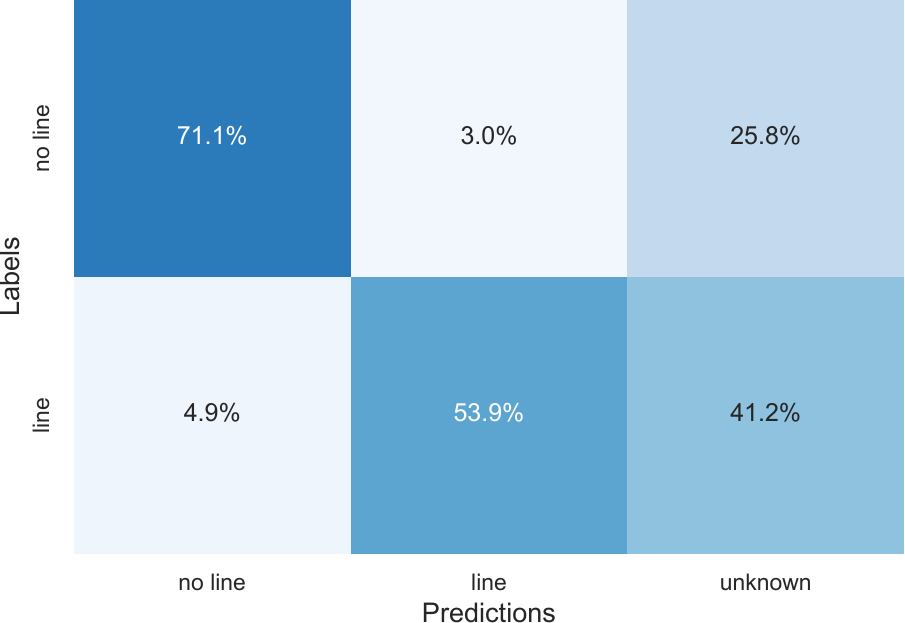}
                \caption{
                    Confusion matrix where every classification with a confidence score below the threshold is considered as ``\emph{unknown}''.
                    Accuracy: \qty{94.6}{\percent}; precision: \qty{93.5}{\percent}; recall: \qty{93.8}{\percent}; F1 score: \qty{93.6}{\percent}; unknown rate: \qty{31.2}{\percent}.
                }
            \end{subfigure}
            \caption[Confusion matrices]{
                Example of a confusion matrix for the \acf*{Si-OG-QD} dataset using a \acf*{CNN} and the cross-validation training method.
                Using the confidence threshold greatly improves the classification performance at the price of \qty{31.2}{\percent} of the patches being labeled as ``\emph{unknown}''.
            }
            \label{fig:confusion-matrices}
        \end{figure}

One can improve the effective model accuracy by excluding model predictions that are below the confidence threshold.
The model outputs can then be interpreted as three-class inferences: ``\emph{line}'' ($\text{output} = 1$ and $\text{confidence} \geq \text{threshold}$), ``\emph{no-line}'' ($\text{output} = 0$ and $\text{confidence} \geq \text{threshold}$), and ``\emph{unknown}'' ($\text{confidence} < \text{threshold}$).
On average, across all the datasets and models, this approach decreased the number of errors by \qty{70}{\percent}, at the price of \qty{11}{\percent} of the patches being classified as ``\emph{unknown}'' (Table~\ref{tab:line-results} and Supplementary Figure~\ref{fig:confusion-matrices}).
This method pushes the \ac{CNN} classification accuracy above \qty{97}{\percent} for all the datasets, which allows for robust autotuning procedures.
However, we did not observe any benefits of using the Bayesian confidence score compared to the classical heuristic.

    \begin{table}[H]
        \centering
        \caption[Transition line detection results]{
        	Line detection results for each dataset and model.
        	The performances are averaged over \num{10} runs using different random seeds.
        	The standard deviation of the run performances represents the variability of the methods.
        	Each run is a cross-validation across every diagram of the dataset.
        	The best test accuracy scores of each dataset are highlighted in bold.
        	The equivalent table without cross-validation is available in Supplementary Table~\ref{tab:line-results-no-cross-validation}.
        }
        \label{tab:line-results}
        \begin{tabular}{|c|c|cc|cc|}
            \hline 
            Dataset &
            Model &
            Accuracy &
            \makecell{Accuracy above\\threshold} &
            \makecell{Error reduction\\using threshold} &
            \makecell{Rate below\\threshold} \\
            \hline 
                                      & BCNN & 96.9\% \std{0.1} & 99.3\%          \std{0.1} & 78.8\% \std{3.0} & 4.8\% \std{0.4}  \\
                                      & CNN  & 96.9\% \std{0.1} & \textbf{99.4\%} \std{0.1} & 82.4\% \std{2.7} & 4.9\% \std{0.5}  \\
            \multirow{-3}{*}{\michel} & FF   & 90.4\% \std{1.1} & 99.3\%          \std{0.1} & 90.8\% \std{1.3} & 22.3\% \std{1.3} \\
            \hline 
                                      & BCNN & 93.3\% \std{0.5} & 96.2\%          \std{0.4} & 48.5\% \std{6.8} & 9.5\% \std{1.1}  \\
                                      & CNN  & 94.6\% \std{0.2} & \textbf{97.4\%} \std{0.2} & 59.6\% \std{4.3} & 8.4\% \std{0.8}  \\
            \multirow{-3}{*}{\louis}  & FF   & 93.1\% \std{0.2} & 96.5\%          \std{0.4} & 57.5\% \std{5.9} & 9.4\% \std{1.6}  \\
            \hline 
                                      & BCNN & 92.5\% \std{0.4} & 96.9\%          \std{0.3} & 56.8\% \std{3.2} & 10.6\% \std{0.7} \\
                                      & CNN  & 90.7\% \std{0.3} & \textbf{98.1\%} \std{0.2} & 77.8\% \std{1.9} & 16.2\% \std{0.5} \\
            \multirow{-3}{*}{\eva}    & FF   & 90.8\% \std{0.2} & 97.9\%          \std{0.1} & 77.8\% \std{1.0} & 17.0\% \std{0.5} \\
            \hline 
        \end{tabular}
       
    \end{table}

    \section{Autotuning}\label{sec:autotuning}

\subsection{Methods}\label{subsec:autotuning-method}

Now that we have established a high-accuracy line detection method using \acp{NN}, we need to define an exploration strategy that uses model classification and uncertainty to efficiently explore the stability diagram space until the region of interest can be located.
One step of this exploration is a cycle of the following: \emph{(i)} patch measurement from a charge sensor on the device (simulated by extracting data from offline diagrams in this study), \emph{(ii)} line detection using the trained \ac{NN}, and \emph{(iii)} deciding the following patch coordinates based on the exploration policy.
The borders of the diagrams constrain the exploration to safe gate voltage ranges.
The number of steps can be seen as a proxy for the relative tuning time, since the measurement is the longest part of the tuning process by orders of magnitude.
The actual duration depends on the number of data points, the sensor type, and the electronic performance.
For example, obtaining an \numproduct{18 x 18} patch of \ac{Si-SG-QD} takes approximately \qty{2}{\minute} by measuring the \ac{SET} current for the \num{324} points over a range of \numproduct{36 x 36} \unit{\mV} using a \emph{Keysight 34465A} multimeter.
Although the measurement time can be shortened using more advanced and integrated equipment, it is likely to remain the limiting factor.
Therefore, our exploration strategy aims to minimize the number of steps while maximizing the tuning success rate.

The typical strategy~\cite{Czischek_2021, Durrer_2020, Lapointe_Major_2020, Ziegler_2023} is to first search for the zero-electron regime, which is characterized by the absence of transition lines in a large area (bottom-left corners in the diagrams shown in Figure~\ref{fig:datasets}).
Then, we count the number of transition lines until we reach the desired regime.
In the case of one-electron tuning, the target location will be between the first and second lines.
However, this simple exploration strategy is very susceptible to misclassification or hardware changes.

\begin{figure}
        \centering
        \includegraphics[width=.85\textwidth]{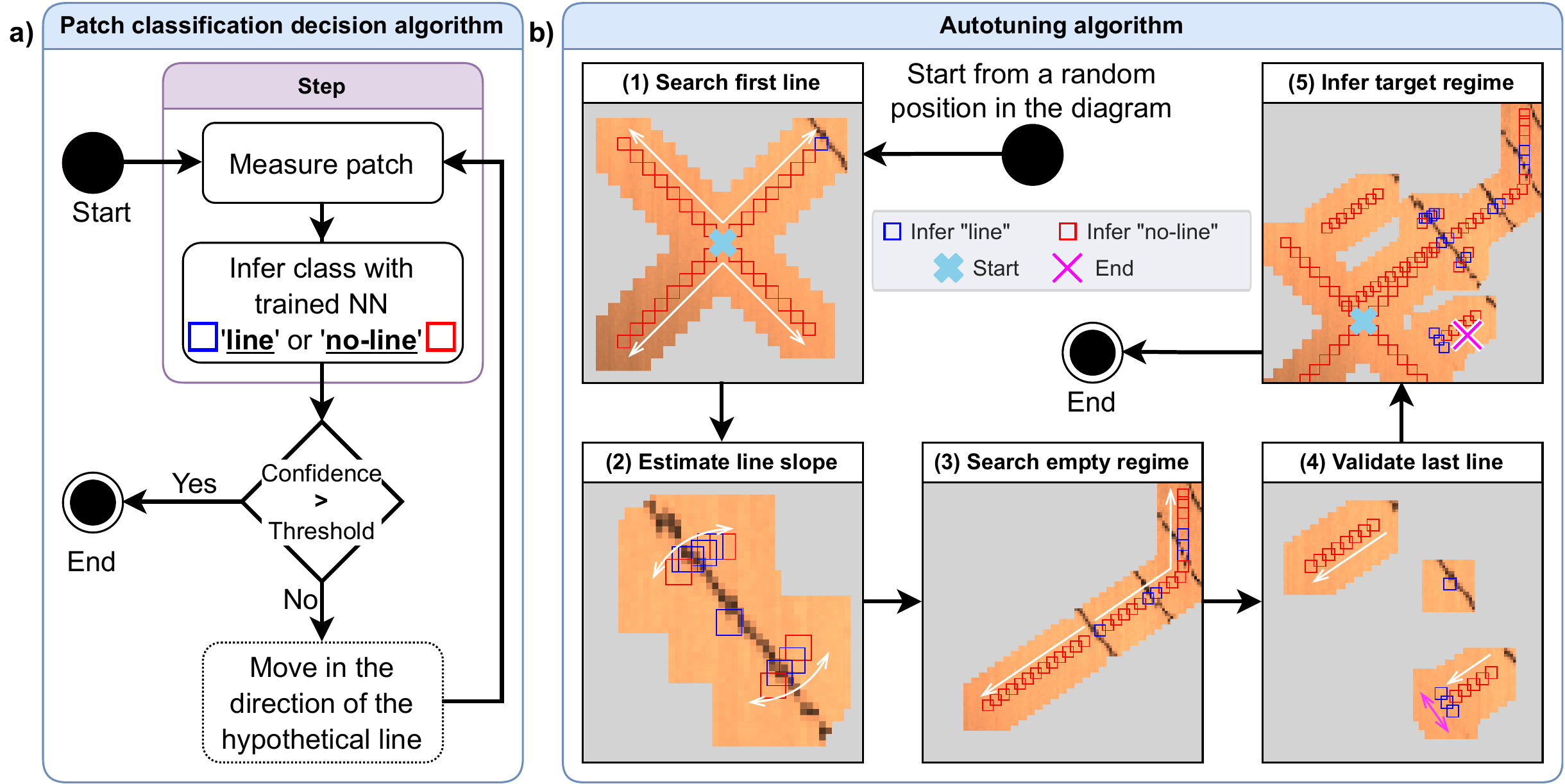}
        \caption[Autotuning algorithm]{
            \textbf{(a)} Schematic representation of the patch classification algorithm.
            The purple box encloses the logic of one exploration step.
            \textbf{(b)} Step-by-step example of the autotuning algorithm using the \ac{GaAs-QD} dataset (complete diagram in Figure~\ref{fig:datasets}b).
            The arrows represent the direction of the exploration, and the gray area is unmeasured space during the current step.
            \emph{(1)} Search the first line by exploring the voltage space in \num{4} directions.
            \emph{(2)} Estimate the line slope by scanning two sequences of patches in circular arcs around the first patch with a line.
            \emph{(3)} Explore the space perpendicularly to the first line to gather information about the distance between lines.
            Stop after reaching \num{3} times the average distance without detecting a new line at lower voltages.
            \emph{(4)} Search for possible missed lines under the first line by scanning multiple sections where we would expect to find a new line according to the average line distance and slope estimation.
            In this example, a fading line is correctly detected at the bottom of the image, but the low confidence of the model triggers a validation procedure.
            More scans are then processed on the hypothetical line direction (purple arrow) until a higher confidence inference validates or invalidates the line's existence.
            \emph{(5)} Deduce the one-electron regime location based on the leftmost line position, the slope, and the average space between lines (all scans from previous steps are represented here).
            The algorithmic details of each step are available in Supplementary Section~\ref{subsec:suppl-exploration-algo}.
        }
        \label{fig:tuning-algo}
    \end{figure}

To make the exploration robust and compatible with most \ac{QD} technologies, we adapt this exploration strategy, as illustrated in Figure~\ref{fig:tuning-algo} and a video\footnote{Method presentation and animated autotuning examples: \href{https://youtu.be/9pPrgrIx9O0}{youtu.be/9pPrgrIx9O0}}.
This approach can adapt on the fly to different line slopes (step 2) and spacings (step 3) while checking for any fading lines (step 4).
When a patch is classified with a confidence score below the threshold, we explore the space surrounding it to reduce the risk of critical failure due to \ac{NN} misclassification.
We validate or refute the presence of a line by taking additional steps in its supposed direction (see the purple arrow in Figure~\ref{fig:tuning-algo}b, step 4) until a patch is classified with high confidence.

We evaluated this exploration strategy for each diagram that contained the one-electron regime within the measured voltage range (nine for each dataset).
We used a k-fold cross-validation method~\cite{Raschka_2018} (Supplementary Figure~\ref{fig:cross-validation}) to test our approach on every valid diagram while preventing the inclusion of testing patches in the training set.
This testing method should be close to an online autotuning situation, where we want to tune a new device based on previous experiments.
The tuning success represents the proportion of final voltage coordinates inside the one-electron regime area out of \qty{50} random starting points for every diagram in the dataset.
The entire experiment is reproduced \qty{10} times using different random seeds that affect the \ac{NN} parameter initialization, the training process, and the random starting points.
In total, \num{1.1e7} steps were evaluated in \qty{81000} offline autotuning simulations using \qty{810} independently trained \acp{NN}.

\subsection{Results}\label{subsec:autotuning-results}

The results for each possible combination of dataset and \ac{NN} are summarized in Table~\ref{tab:results}, showing the autotuning success rate with and without leveraging the \ac{NN}'s confidence score (uncertainty-based).
Using the model uncertainty information consistently reduces the number of tuning failures (\qty{-53}{\percent} on average) at the price of a few additional steps (\qty[retain-explicit-plus]{+22}{\percent} on average).
The best improvement is observed for the \ac{CNN} on \ac{Si-SG-QD}, where the tuning success rate increased from \qty{88.8}{\percent} to \qty{99.5}{\percent} when the autotuning procedure exploited the confidence score (\qty{-95}{\percent} tuning failures and \qty[retain-explicit-plus]{+8.5}{\percent} steps).

Surprisingly, the confidence score provided by the Bayesian version of the \ac{CNN} does not provide a significant improvement over the simple heuristic obtained using the standard \ac{CNN}.
The average line detection accuracy is similar for the \ac{CNN} and \ac{BCNN} on the \ac{Si-SG-QD} dataset, and the line detection performance is slightly better for the \ac{BCNN} on \ac{Si-OG-QD}.
However, on every dataset, the \ac{CNN} reaches a higher tuning success by using the confidence heuristic score (highlighted in bold in Table~\ref{tab:results}).
Even in the case where the \ac{BCNN} line detection accuracy is higher (\ac{Si-OG-QD}), the tuning success benefits more from the \ac{CNN} confidence score (\qty[retain-explicit-plus]{+16.4}{\percent} tuning success) than the one from the \ac{BCNN} (\qty[retain-explicit-plus]{+8.9}{\percent} tuning success).
This result suggests that the number of critical misclassifications above the confidence threshold is more frequent with the Bayesian \ac{NN}'s uncertainty compared to the classical confidence heuristic score.

The line detection accuracy can provide valuable indications regarding the \ac{NN}'s performance, but it is not entirely correlated to the tuning success rate, since some misclassifications are much more harmful than others during diagram exploration.
If a classification error occurs near a line (soft errors in Figure~\ref{fig:full-map}c), it will likely be corrected at the next step.
However, an error on a line, or far from it, will likely affect the final tuning result.
For example, if a noisy patch is wrongly classified as a line in the empty regime area, the number of charge carriers inside the \ac{QD} will be overestimated by one.

The uncertainty-based autotuning method with convolution models allows for a nearly perfect tuning success rate (\textgreater\qty{99}{\percent}) on the \ac{Si-SG-QD} dataset.
This high performance is attributable to \emph{(i)} the convolution layers' ability to efficiently filter the oscillating background noise, \emph{(ii)} the exploitation of the model confidence score to prevent most critical errors, and \emph{(iii)} the straight transition lines that simplify the exploration.
The line detection task also benefits from the large training set (\num{72000} patches) and the low variability between diagram features.

The \ac{GaAs-QD} and \ac{Si-OG-QD} datasets present more feature diversity in their diagrams with a smaller training set, which makes the classification task more challenging.
The shape of the transition lines is also less predictable than the \ac{Si-SG-QD} dataset, which increases the risk of missing a line or failing to detect its slope and spacing.
The significantly higher tuning performance without cross-validation (Supplementary Table~\ref{tab:results-no-cross-validation}) suggests that the training sets do not cover the diversity of the stability diagrams.
This is well illustrated by the \acp{CNN} on \ac{GaAs-QD}, where the tuning success rate is improved by \qty{11.6}{\percent} without cross-validation (\qty{92.2}{\percent} versus \qty{80.6}{\percent}).
This hypothesis is confirmed by the high variability of the tuning success rates between the diagrams in these two datasets.
The results are very polarized between the diagrams for which the one-electron regime is found nearly \qty{100}{\percent} of the time, while some are more complex to tune and exhibit a very low success rate. 
A larger or more consistent dataset could reduce the tuning success loss observed with cross-validation.

Overall, noise and hardware imperfections are relatively easy to manage for the line detection task but more challenging to deal with at the diagram scale during exploration.
Therefore, the reliability of the autotuning procedure is mainly affected by the shape and consistency of the transition lines.

    \begin{table}[H]
        \centering
        \rowcolors{2}{}{lightgray!30}  
        \caption{
        	Autotuning results for each dataset and model, with and without using the model uncertainty information provided by the confidence score.
        	The line detection accuracy and tuning success variability are computed over \num{10} runs using different random seeds.
        	Each run is a cross-validation over every diagram in the dataset.
        	The best tuning success rates for each dataset are highlighted in bold.
        	The tuning success rates can be compared to the baselines presented in Supplementary Table~\ref{tab:baselines}.
        	The equivalent table without cross-validation is available in Supplementary Table~\ref{tab:results-no-cross-validation}.
        }
        \label{tab:results}
        \begin{tabular}{|>{\cellcolor{white}}c|>{\cellcolor{white}}c|>{\cellcolor{white}}c|ccc|}
            \hline 
            Dataset &
            Model &
            \makecell{Line detection\\accuracy} &
            \makecell{Uncertainty-\\based tuning} &
            \makecell{Average step\\number} &
            Tuning success \\
            \hline 
                                      &                        &                                    & Yes & 164 & 99.2\%          \std{0.7} \\
                                      & \multirow{-2}{*}{BCNN} & \multirow{-2}{*}{96.9\% \std{0.1}} & No  & 148 & 88.2\%          \std{2.3} \\
                                      &                        &                                    & Yes & 165 & \textbf{99.5\%} \std{0.7} \\
                                      & \multirow{-2}{*}{CNN}  & \multirow{-2}{*}{96.9\% \std{0.1}} & No  & 152 & 88.8\%          \std{2.4} \\
                                      &                        &                                    & Yes & 194 & 72.4\%          \std{6.3} \\
            \multirow{-6}{*}{\michel} & \multirow{-2}{*}{FF}   & \multirow{-2}{*}{90.4\% \std{1.1}} & No  & 122 & 23.9\%          \std{5.4} \\
            \hline 
                                      &                        &                                    & Yes & 104 & 75.3\%          \std{4.2} \\
                                      & \multirow{-2}{*}{BCNN} & \multirow{-2}{*}{93.3\% \std{0.5}} & No  &  92 & 55.9\%          \std{3.8} \\
                                      &                        &                                    & Yes & 103 & \textbf{80.6\%} \std{3.9} \\
                                      & \multirow{-2}{*}{CNN}  & \multirow{-2}{*}{94.6\% \std{0.2}} & No  &  93 & 72.4\%          \std{2.9} \\
                                      &                        &                                    & Yes & 105 & 72.8\%          \std{4.5} \\
            \multirow{-6}{*}{\louis}  & \multirow{-2}{*}{FF}   & \multirow{-2}{*}{93.1\% \std{0.2}} & No  &  92 & 58.4\%          \std{3.4} \\
            \hline 
                                      &                        &                                    & Yes & 185 & 75.2\%          \std{2.5} \\
                                      & \multirow{-2}{*}{BCNN} & \multirow{-2}{*}{92.5\% \std{0.4}} & No  & 155 & 66.3\%          \std{3.4} \\
                                      &                        &                                    & Yes & 193 & \textbf{78.1\%} \std{1.7} \\
                                      & \multirow{-2}{*}{CNN}  & \multirow{-2}{*}{90.7\% \std{0.3}} & No  & 150 & 61.7\%          \std{3.0} \\
                                      &                        &                                    & Yes & 200 & 78.0\%          \std{2.5} \\
            \multirow{-6}{*}{\eva}    & \multirow{-2}{*}{FF}   & \multirow{-2}{*}{90.8\% \std{0.2}} & No  & 151 & 59.3\%          \std{2.4} \\
            \hline 
        \end{tabular}
    \end{table}

    \section{Discussion}\label{sec:discussion}

The proposed method can be used to automate the charge tuning of any spin-based semiconductor single \ac{QD} using less than \num{10} annotated stability diagrams for training and some basic prior knowledge of the device characteristics (expected working range voltage, approximate transition lines slope, and spacing).
The autotuning procedure is resilient to noise and physical imperfections but sensitive to high device-to-device variability and training dataset quality (number of diagrams and feature coverage).
This is well illustrated by the \ac{Si-SG-QD} dataset, which features strong noise but consistent diagrams, allowing for a robust autotuning procedure with only \num{23} tuning failures over \num{4500} \ac{CNN} uncertainty-based tuning simulations.
Therefore, this method, associated with a good fabrication yield, has the potential to enable parallel charge tuning of large \ac{QD} arrays, which is currently not practical using a manual approach.

This study demonstrates the benefits of using \ac{NN} uncertainty to automate the tuning of \acp{QD} based on only partial measurements of the stability diagrams.
The confidence score provides valuable information regarding the model prediction, which can be used to design robust autotuning procedures.
However, we did not see a clear benefit of using a Bayesian \ac{NN} over a standard one, even though Bayesian models are specifically designed to provide uncertainty measurements.
This behavior could be explained by the distributional uncertainty being poorly captured by the Bayesian \acp{NN}~\cite{Izmailov_2021}, which could be the predominant source of uncertainty when using the cross-validation testing method.
Moreover, many approximations~\cite{Goan_2020, Graves_2011} are necessary to keep the Bayesian \ac{NN} training and inference tractable, which could be detrimental to the overall quality of the model and its coherence with the initial Bayesian framework.
Thus, the additional complexity and computing cost induced by the Bayesian \ac{NN} is hard to justify when a simpler \ac{NN} can provide a more reliable confidence score.
In future work, other methods of uncertainty quantification~\cite{Abdar_2021} (e.g., dropout as Bayesian approximation~\cite{Gal_2016} or ensemble learning~\cite{Lakshminarayanan_2016}) and prior selection~\cite{Fortuin_2021, Silvestro_2020} could be evaluated to improve the reliability of the confidence score.
Bayesian optimization associated with classical machine learning~\cite{Gebhart_2023, Szulakowska_2022} also appears to be a promising avenue to harness the complexity of quantum engineering.

While applying quantum computing gates requires more than one \ac{QD}, we developed and tested our approach on single-\ac{QD} datasets as a first step toward robust double- and triple-\ac{QD} autotuning.
Since \acp{NN} are known for their good scalability, we expect this method to be expandable to larger input dimensions~\cite{Krause_2022, Moon_2020} and greater numbers of classes.
When the problem's complexity increases, the benefits of the confidence score could become even more important.

To avoid any wiring bottleneck~\cite{Reilly_2019} between the fridge and external electronics, this calibration method could be integrated near the \acp{QD} in the \qty{4}{\K} stage of the cryogenic environment.
One way to meet this requirement would be to transfer the trained \ac{NN} to low-power and cryo-compatible hardware, such as circuits based on arrays of memristive devices~\cite{Mouny_2023, Marcotte_2023, Dawant_2024}.
This specialized hardware takes advantage of in-memory computing~\cite{Christensen_2022, Wang_2020} to efficiently perform the multiply--accumulate operations~\cite{Amirsoleimani_2020} required for \ac{NN} inference.
Therefore, memristor-based systems represent promising candidates to realize scalable autotuning from inside the fridge with close-loop measurements and minimal disturbance over the \acp{QD}.
Future studies should be performed to simulate these hardware implementations and demonstrate real-time online autotuning using the proposed method.


    \section*{Acknowledgments}\label{sec:acknowledgments}

    V.Y. acknowledges Stefanie Czischek, who inspired this study by working on a first version of the autotuning procedure.
    The authors also acknowledge the experimentalists who provided the stability diagram measurements used in this paper
    (Michel Pioro-Ladrière, Marc-Antoine Roux, Marc-Antoine Genest, Julien Camirand-Lemire, and Sophie Rochette).

    V.Y., B.G., Y.B., and D.D. acknowledge support from the National Science Engineering Research Council of Canada, Grant ALLRP 580722–22, and the Fonds de Recherche du Québec---Nature et Technologies, Grant 300253.

    C.R., J.R., A.M., D.L., and E.D.F acknowledge support from the FRQNT établissement de la relève professorale, Grant 2020--NC--268397, and the CRSNG, Grant RGPIN--2020--0573.

    R.G.M. acknowledges support from NSERC and the Perimeter Institute for Theoretical Physics.
    Research at the Perimeter Institute is supported in part by the Government of Canada through the Department of Innovation, Science and Economic Development Canada and by the Province of Ontario through the Ministry of Economic Development, Job Creation and Trade.

    \section*{Conflicts of interest}\label{sec:conflict}

    The authors declare no conflicts of interest.

    \section*{Author contributions}\label{sec:contributions}

    All authors contributed to this article and approved of the submitted version.

    Victor Yon: methodology, datasets aggregation and processing, software implementation, running experiments, results analysis and visualization, writing---original draft preparation.

    Bastien Galaup: contributed to software implementation, running experiments, review and editing.

    Claude Rohrbacher and Joffrey Rivard: provided data for the \ac{Si-OG-QD} dataset, provided expertise and technical assistance with quantum dot technology, review and editing.

    Clément Godfrin, Ruoyu Li, Stefan Kubicek, and Kristiaan De Greve: manufactured the devices used to obtain the \ac{Si-OG-QD} dataset.

    Louis Gaudreau: provided data for the \ac{GaAs-QD} dataset, provided expertise and technical assistance on quantum dot technology, review and editing.

    Yann Beilliard: methodology, supervision, funding acquisition, review and editing.

    Eva Dupont-Ferrier and Roger Melko: supervision, review and editing.

    Dominique Drouin: methodology, supervision, funding acquisition, review and editing.

    \section*{Code and Data Availability}\label{sec:data}

    The Python source code used to aggregate and build the quantum dot stability diagram dataset is publicly accessible on \emph{GitHub}: \href{https://github.com/3it-inpaqt/qdsd-dataset}{github.com/3it-inpaqt/qdsd-dataset}.

    The Python source code used to run all the experiments presented in this article is publicly accessible on \emph{GitHub}: \href{https://github.com/3it-inpaqt/dot-calibration-v2/tree/offline-article}{github.com/3it-inpaqt/dot-calibration-v2/tree/offline-article}.

    The raw and processed stability diagram measurements used to train and test the model presented in this article are publicly available for download from \citet{qdsd}.

    All simulation outputs and results used to generate the tables and figures presented in this article are publicly available for download from \citet{run_outputs}.

    \newpage
    \bibliographystyle{unsrtnat}
    \bibliography{references}

    \beginsupplement

    \begin{center}
        \textbf{\huge Supplementary Materials: Robust Quantum Dots Charge Autotuning using Neural Network Uncertainty}
    \end{center}

    \section{Datasets}\label{sec:suppl-datasets}

    \begin{table}[H]
        \centering
          \caption[Summary of dataset specifications]
        {Summary of dataset specifications.
        	\acs{SET} and \acs{QPC} stand for \acl{SET} and \acl{QPC}, respectively.
        }
        \label{tab:dataset-specs}
        \begin{tabular}{|c|c|c|c|}
            \hline
            \textbf{Dataset short name}                                      & \textbf{\michel}      & \textbf{\louis}  & \textbf{\eva}   \\ \hline
            \textbf{Dataset full name} &
                \makecell{Silicon split gate\\quantum dot} &
                \makecell{Gallium arsenide\\quantum dot} &
                \makecell{Silicon overlapping\\gate quantum dot} \\ \hline
            \textbf{Current measurement method}                              & \acs{SET}             & \acs{QPC}           & \acs{SET}                 \\ \hline
            \textbf{Pixel size}                                              & \qty{1}{\mV}          & \qty{2.5}{\mV}      & \qty{2}{\mV}              \\ \hline
            \textbf{Detection area offset}                                   & \qty{6}{pixels}       & \qty{7}{pixels}     & \qty{6}{pixels}           \\ \hline
            \textbf{Prior line distance}                                     & \qty{30}{\mV}         & \qty{16}{\mV}       & \qty{30}{\mV}             \\ \hline
            \makecell{\textbf{Prior line slope}\\
                (\ang{0} = horizontal, \ang{90} = vertical)}                 & \ang{75}              & \ang{45}            & \ang{-10}                 \\ \hline
            \textbf{Use last-line validation}                                & No (not necessary)    & Yes                 & Yes                       \\ \hline
            \makecell{\textbf{Number of diagrams with}\\
                \textbf{transition line annotations}\\
                (used for line detection training and test)}                 & \num{17}              & \num{9}             & \num{12}                  \\ \hline
            \makecell{\textbf{Number of diagrams with}\\
                \textbf{transition line and charge}\\
                \textbf{area annotations}\\
                (used for autotuning test)}                                  & \num{9}               & \num{9}             & \num{9}                   \\ \hline
            \makecell{\textbf{Number of patches}\\
                ($18\times18$ pixels each)}                                  & \num{72182}           & \num{4349}          & \num{48081}               \\ \hline
            \makecell{\textbf{Patch class ratio}\\
                (no-line / line)}                                            & \num{33.6}            & \num{3.3}           & \num{8.2}                 \\ \hline
            \textbf{References}                                              &\cite{Rochette_2019}   &\cite{Gaudreau_2009} &\cite{Elsayed_2024}        \\ \hline
        \end{tabular}
    \end{table}

    \subsection{Data selection}\label{subsec:suppl-data-selection}

    The stability diagrams used in this article are the result of experimental measurements performed in prior studies on \ac{QD} devices~\cite{Rochette_2019, Gaudreau_2009}.  
    Thus, most of the measurements were not suitable for this work, and a selection was made.
    Only stability diagrams meeting the following criteria were included in the final datasets:

    General criteria:
    \begin{itemize}
        \item All diagrams of the same dataset are measured using similar devices.
        Hardware differences are acceptable between datasets, but consistency is required for each independent training.
        \item A similar measurement step size is required between G1 and G2.
        A step size missmatch between the two measured gates makes pixel interpolation less reliable.
        \item Double-\ac{QD} diagrams are excluded.
    \end{itemize}

    Transition line detection patch datasets:
    \begin{itemize}
        \item At least one transition line must be visible or partially visible in the diagram.
    \end{itemize}

    Autotuning datasets:
    \begin{itemize}
        \item At least three transition lines must be visible or partially visible in the diagram.
        \item A clearly identifiable empty regime must be in the measured range (no line for at least two times the average space between visible lines).
    \end{itemize}

    \subsection{Data processing}\label{subsec:suppl-data-processing}

    The long-term goal of this project is to implement an automated control loop as close as possible to the quantum device.
    Thus, we want to minimize the preprocessing of the input data to avoid any computation overhead inside the cryogenic environment.
    This is why we do not compute the derivative on the patches or perform complex data processing, such as the method proposed by \citet{Czischek_2021}.
    Furthermore, most statistical methods require a large sample size to be reliable, which is incompatible with our small patch measurement strategy.

    To improve the convergence of the models, we normalized the input model to a range between \num{0} and \num{1} based on the smallest and largest current value of each patch.

    \subsection{Diagram annotation}\label{subsec:suppl-annotation}

    We manually annotated the transition lines and the charge regime areas using \emph{Labelbox}~\cite{Labelbox} at the diagram level.
    When the transition lines were mixed with background noise, we used the derivative of the images to place more accurate annotations.
    We only annotated the lines that we could visually identify.
    We never filled the gap between two sections of a line, even when it was clear that they were a continuation of the same transition.

    The voltage space was segmented into charge regime areas annotated from \num{0} to \num{3} electrons, then the voltage space corresponding to more than \num{3} electrons was annotated as ``\emph{4+}''.
    The location of each area was deduced from the position of the transition lines and the knowledge that the empty regime was located at the bottom left of the images.
    When the regime of a section was ambiguous or difficult to identify (too noisy, fading line, few pixels around a line), no annotations were set, and the regime of this voltage space was considered as ``\emph{unknown}'' during the experiment (examples in Figure~\ref{fig:datasets}e,f).
    When an autotuning ended in an ``\emph{unknown}'' regime, it was always considered as a failure.

    Annotating the transition lines and charge regimes of one stability diagram took between \num{5} and \num{30} min, depending on their complexity.
    It took approximately \qty{10}{\hour} to annotate the \num{38} stability diagrams used in this study.

    \subsection{Patch segmentation and automatic class labeling}\label{subsec:patch-segmentation}

    The patch size was fixed at \numproduct{18 x 18} pixels as a tradeoff between measurement time and classifier performance.
    With a smaller patch size, it becomes difficult to distinguish a transition line from noise.
    A larger patch would directly increase the time required to measure it, reducing the exploration efficiency.

    To train and test the \acp{NN}, we split the diagrams into patches by sliding an \numproduct{18 x 18} window with \num{10} overlapping pixels between each step.
    We then automatically assigned a class to each patch using the manual transition line annotations set previously (see Section~\ref{subsec:suppl-annotation}).
    If at least one line intersected with the detection area (Figure~\ref{fig:network}b), the patch was categorized as ``\emph{line}''; otherwise, it was categorized as ``\emph{no-line}''.

    \subsection{Subset splitting}\label{subsec:dataset-splitting}

    Each patch dataset was split into three mutually exclusive subsets: training, validation, and test.

\begin{figure}[H]
            \centering
            \begin{subfigure}{0.45\textwidth}
                \centering
                \includegraphics[width=\textwidth]{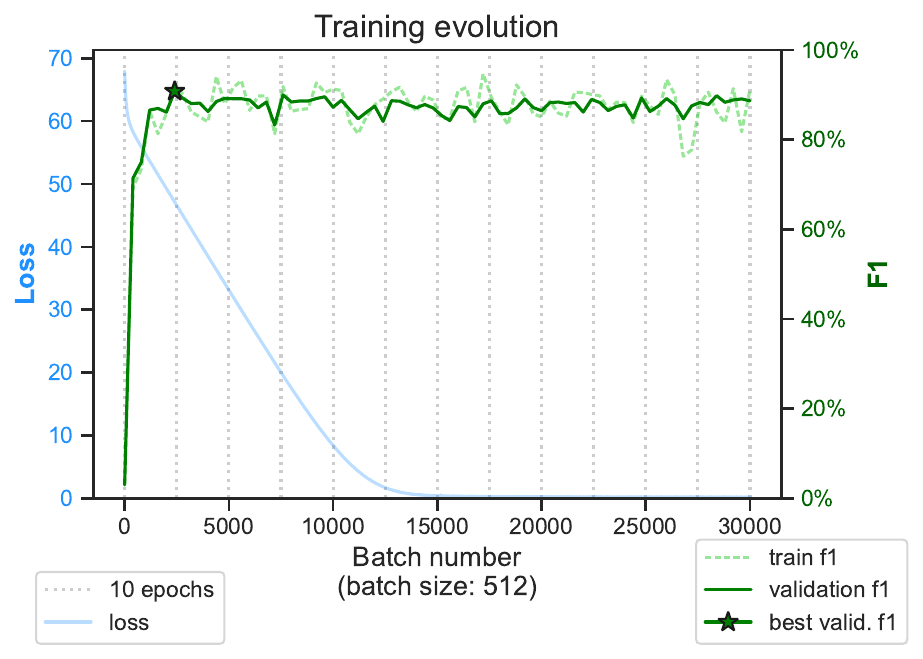}
                \caption{\Acf*{BCNN}}
            \end{subfigure}
            \begin{subfigure}{0.45\textwidth}
                \centering
                \includegraphics[width=\textwidth]{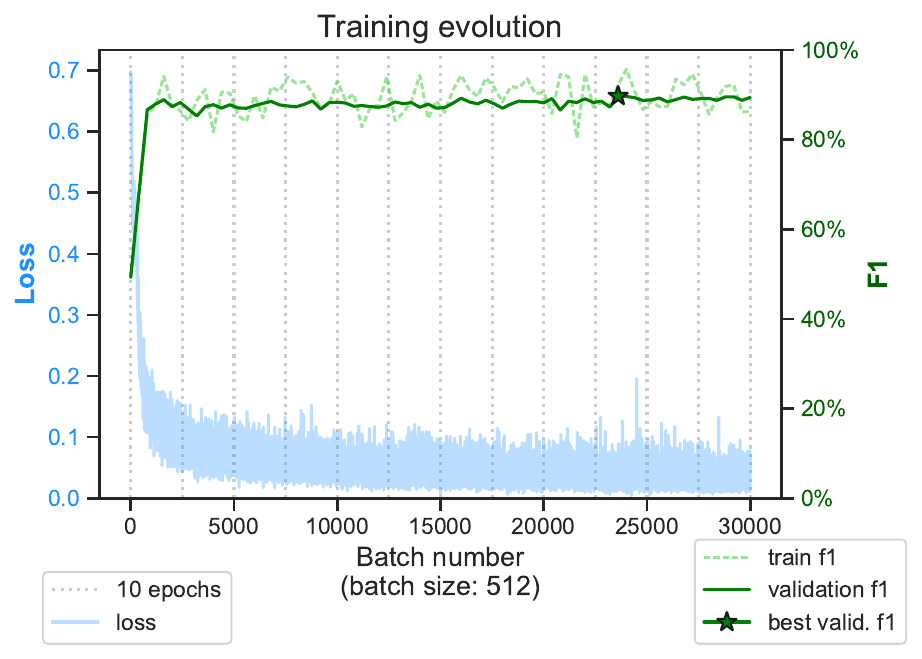}
                \caption{\Acf*{CNN}}
            \end{subfigure}
            \begin{subfigure}{0.45\textwidth}
                \centering
                \includegraphics[width=\textwidth]{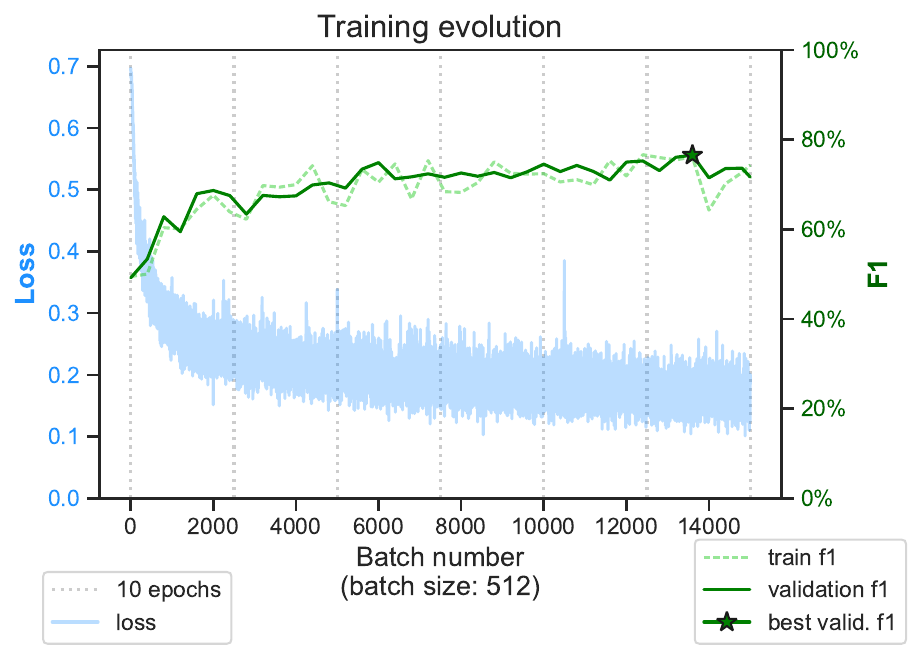}
                \caption{\Acf*{FF}}
            \end{subfigure}
            \caption[Training progress examples]{
                Examples of training progress for the \acf*{Si-SG-QD} dataset using the different models and the cross-validation method.
                The F1 score is the harmonic mean of the precision and recall.
            }
            \label{fig:train-progress}
        \end{figure}

    Each subset was used for a specific purpose:
    \begin{description}[labelindent=1.5cm, leftmargin=2cm, rightmargin=2cm]
        \item[Training set:] used to train the line detection model.
        \item[Validation set:] used after training to select the step corresponding to the best model (green stars in Figures~\ref{fig:train-progress}) and to calibrate the confidence thresholds (Figures~\ref{fig:threshold-calibration}).
        \item[Test set:] used to evaluate the line classification accuracy and the autotuning success rate after training and calibration.
    \end{description}

    In the case of training with cross-validation (Figure~\ref{fig:cross-validation}), the subsets were split as follows:
    \begin{description}[labelindent=1.5cm, leftmargin=2cm, rightmargin=2cm]
        \item[Training set:] composed of \qty{90}{\percent} of the patches, randomly selected among the training diagrams.
        \item[Validation set:] composed of \qty{10}{\percent} of the patches, randomly selected among the training diagrams.
        \item[Test set:] composed of every patch of the testing diagram.
    \end{description}

    In the case of training without cross-validation, the subsets were split as follows:
    \begin{description}[labelindent=1.5cm, leftmargin=2cm, rightmargin=2cm]
        \item[Training set:] composed of \qty{70}{\percent} of the patches, randomly selected among every diagram.
        \item[Validation set:] composed of \qty{10}{\percent} of the patches, randomly selected among every diagram.
        \item[Test set:] composed of \qty{20}{\percent} of the patches, randomly selected among every diagram.
    \end{description}

    \subsection{Stability diagram samples}\label{subsec:digram-samples}

\begin{figure}[H]
        \centering
        \begin{subfigure}{0.45\textwidth}
            \centering
            \includegraphics[width=\textwidth]{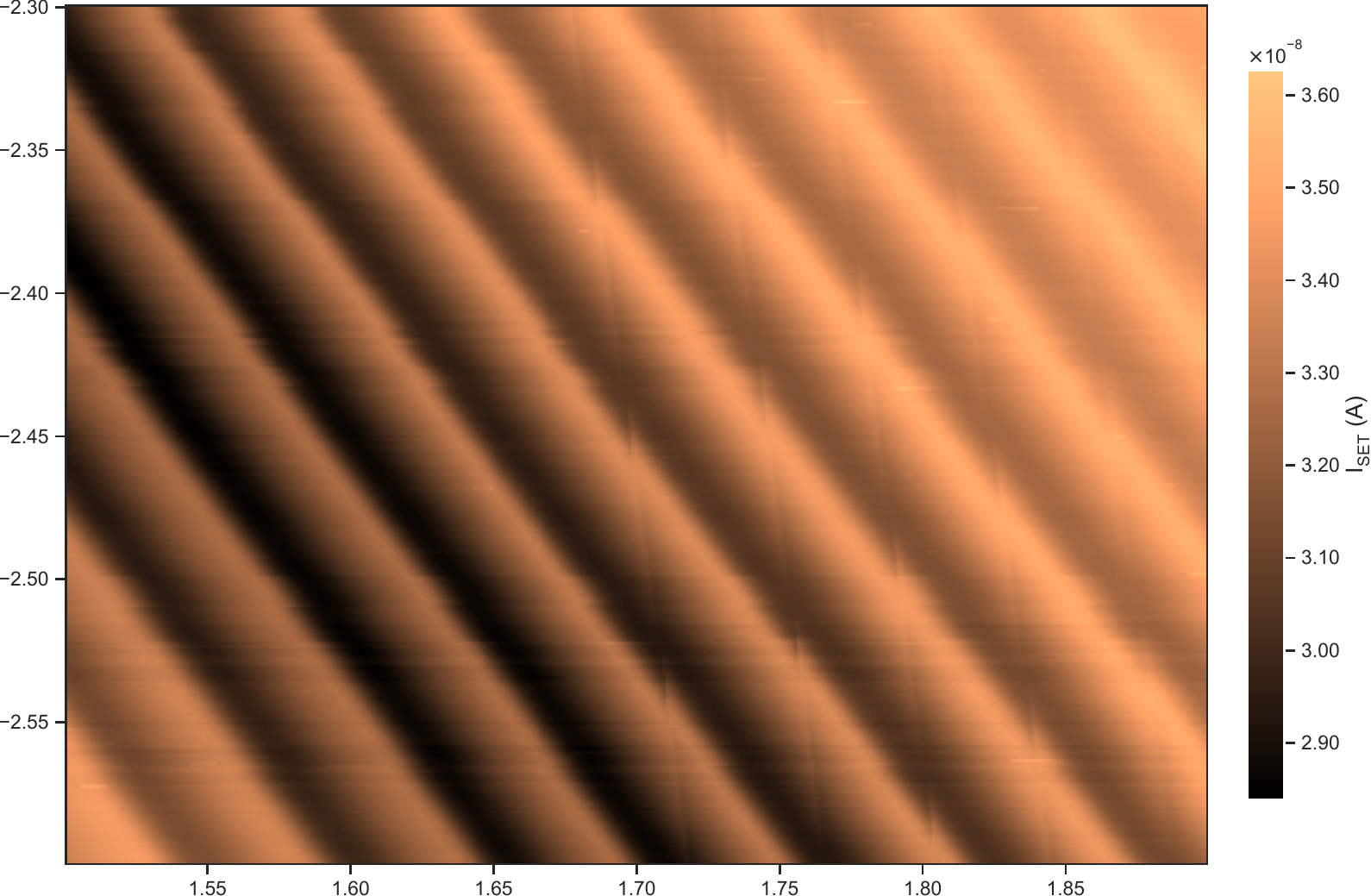}
        \end{subfigure}
        \begin{subfigure}{0.45\textwidth}
            \centering
            \includegraphics[width=\textwidth]{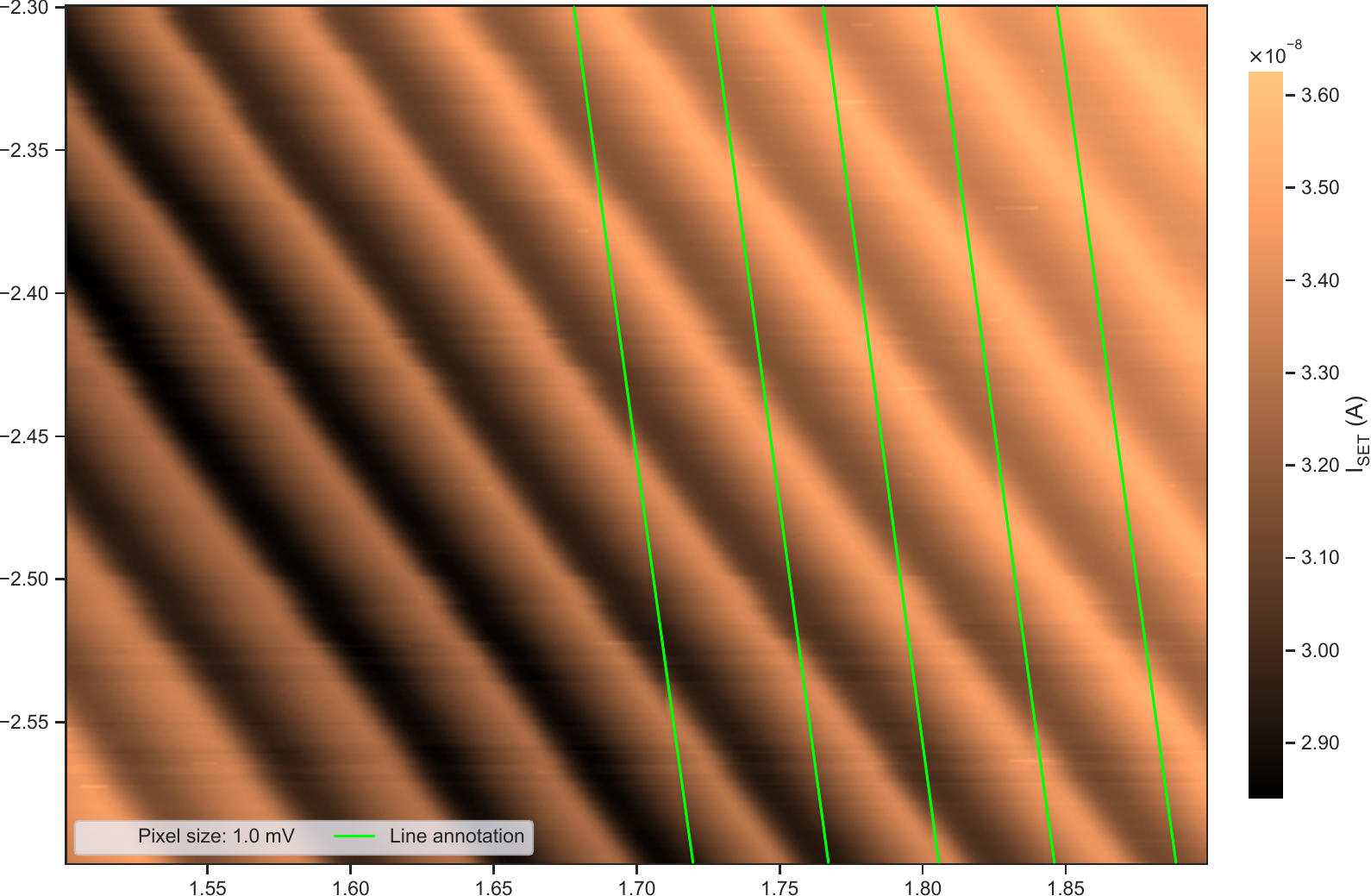}
        \end{subfigure}
        \begin{subfigure}{0.90\textwidth}
            \centering
            \includegraphics[width=\textwidth]{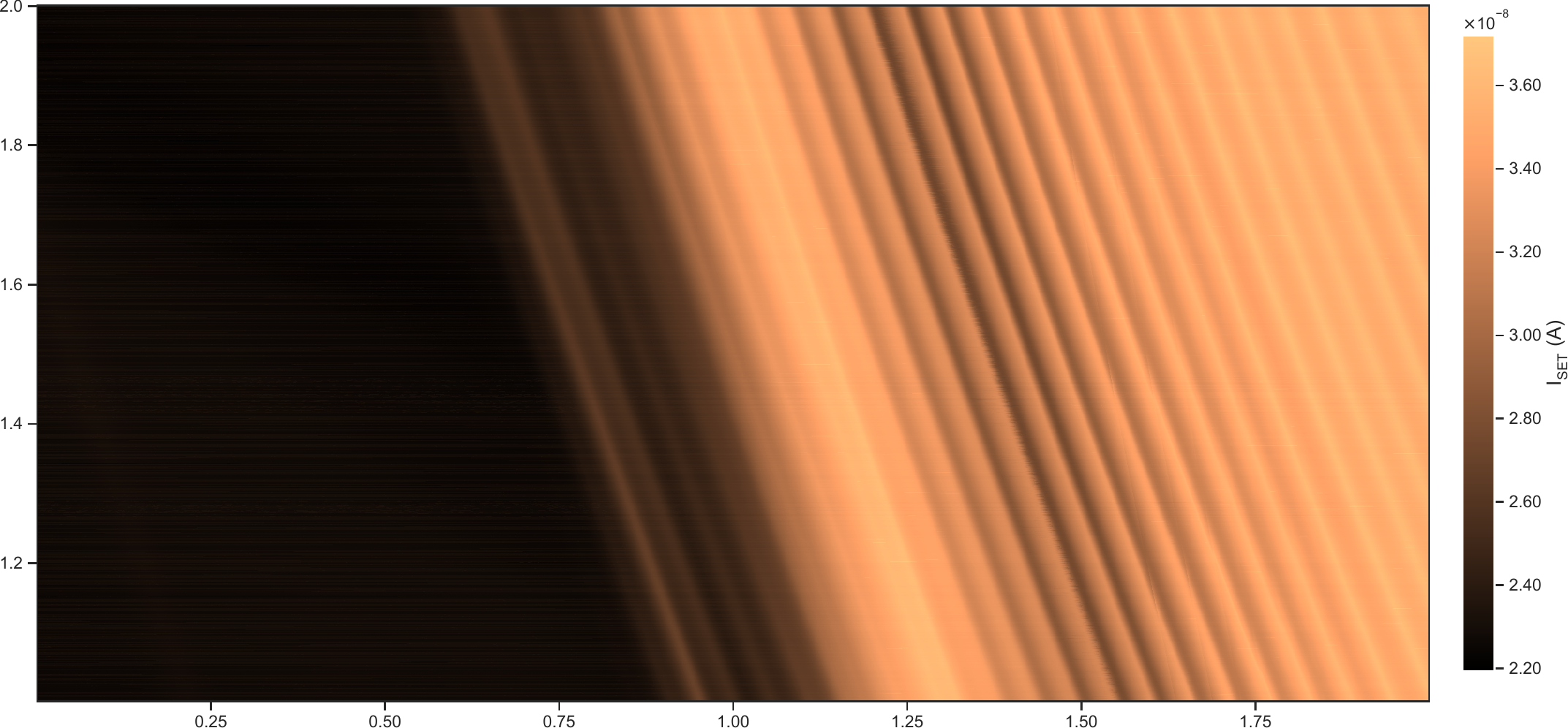}
        \end{subfigure}
        \begin{subfigure}{0.90\textwidth}
            \centering
            \includegraphics[width=\textwidth]{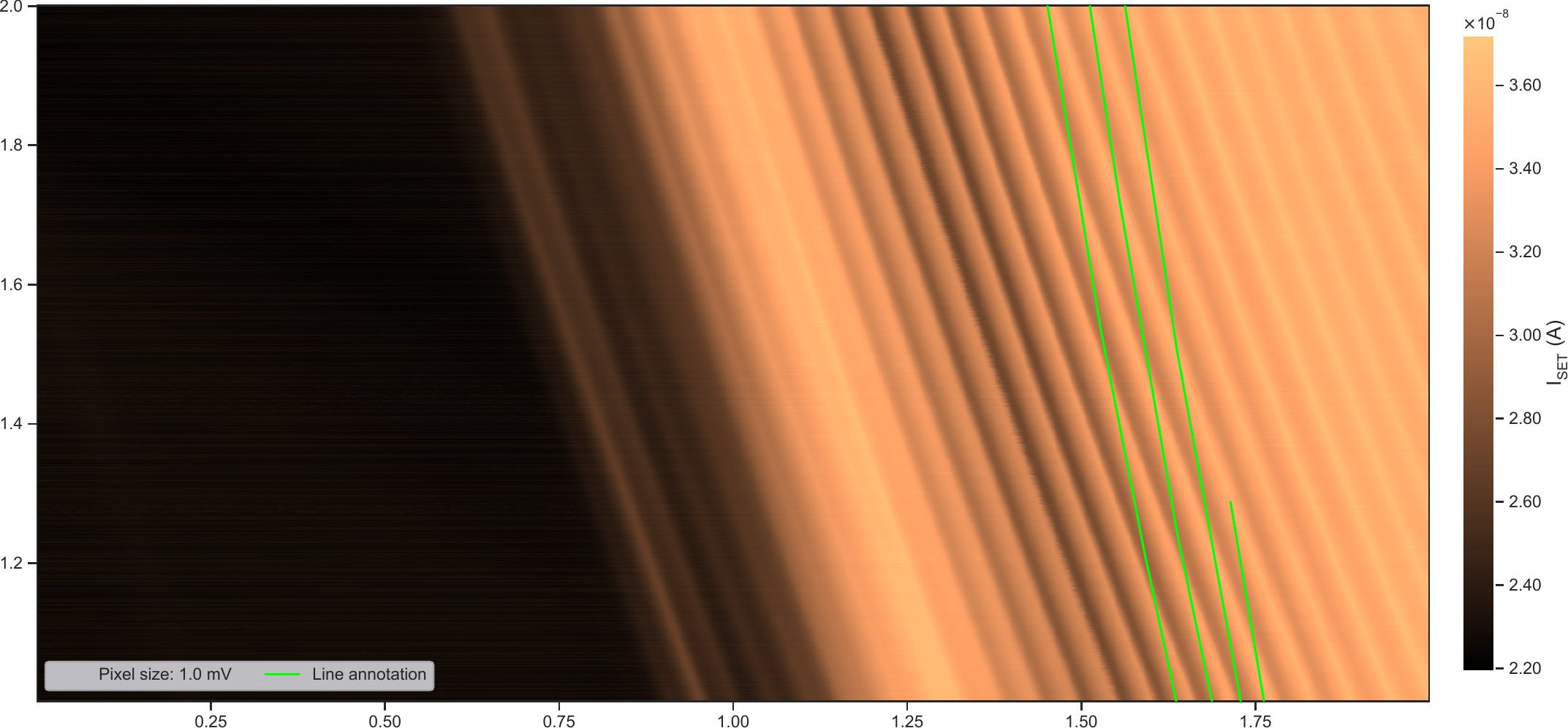}
        \end{subfigure}
        \caption[\acs*{Si-SG-QD} stability diagram samples]{
            Two \acf*{Si-SG-QD} stability diagram examples, without and with transition line annotations.
            The $x$-axes correspond to the G1 gate voltage, and the $y$-axes correspond to the G2 gate voltage.
        }
        \label{fig:digram-samples-michel}
    \end{figure}

\begin{figure}[H]
        \centering
        \begin{subfigure}{0.45\textwidth}
            \centering
            \includegraphics[width=\textwidth]{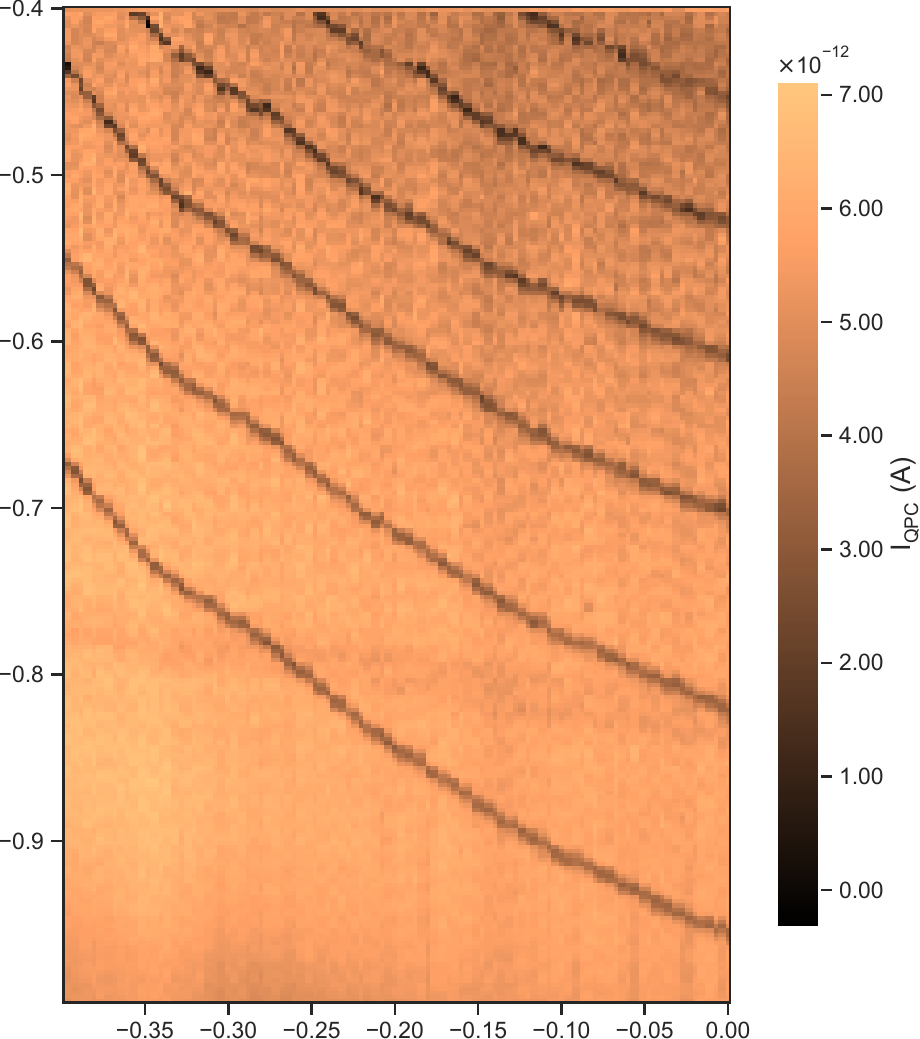}
        \end{subfigure}
        \begin{subfigure}{0.45\textwidth}
            \centering
            \includegraphics[width=\textwidth]{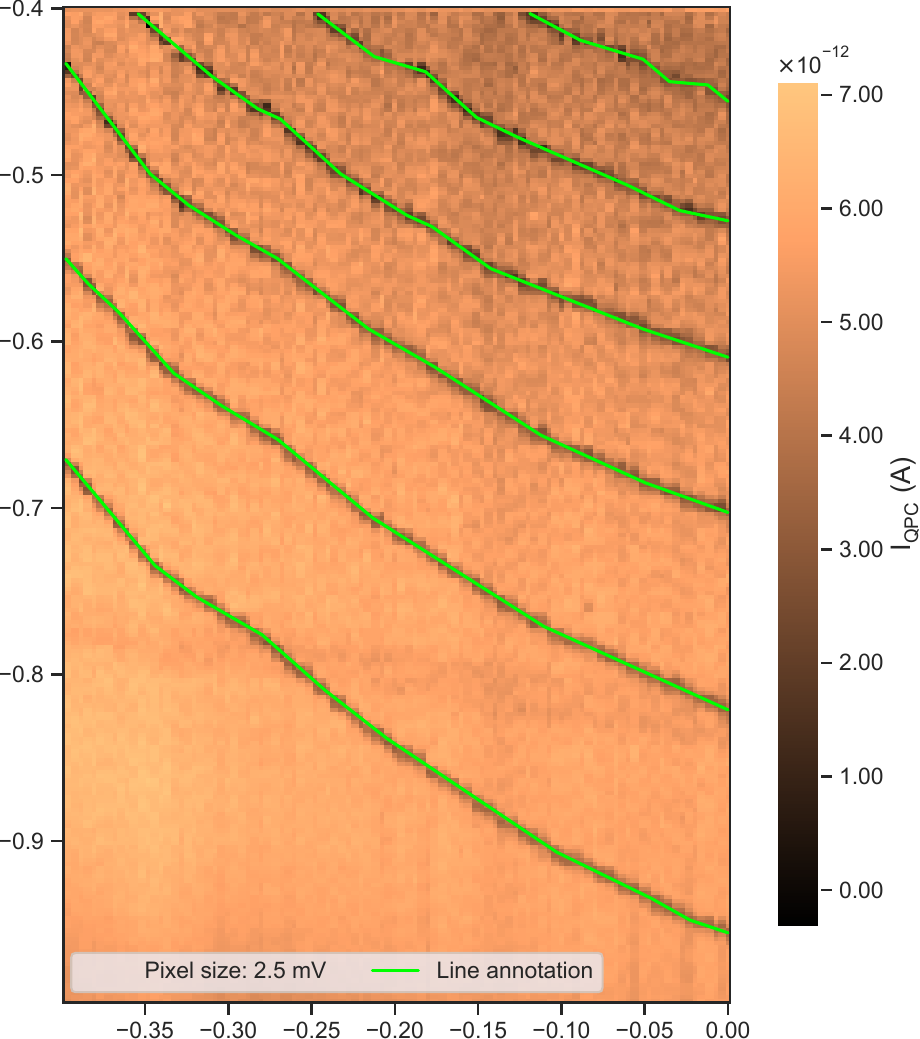}
        \end{subfigure}
        \begin{subfigure}{0.45\textwidth}
            \centering
            \includegraphics[width=\textwidth]{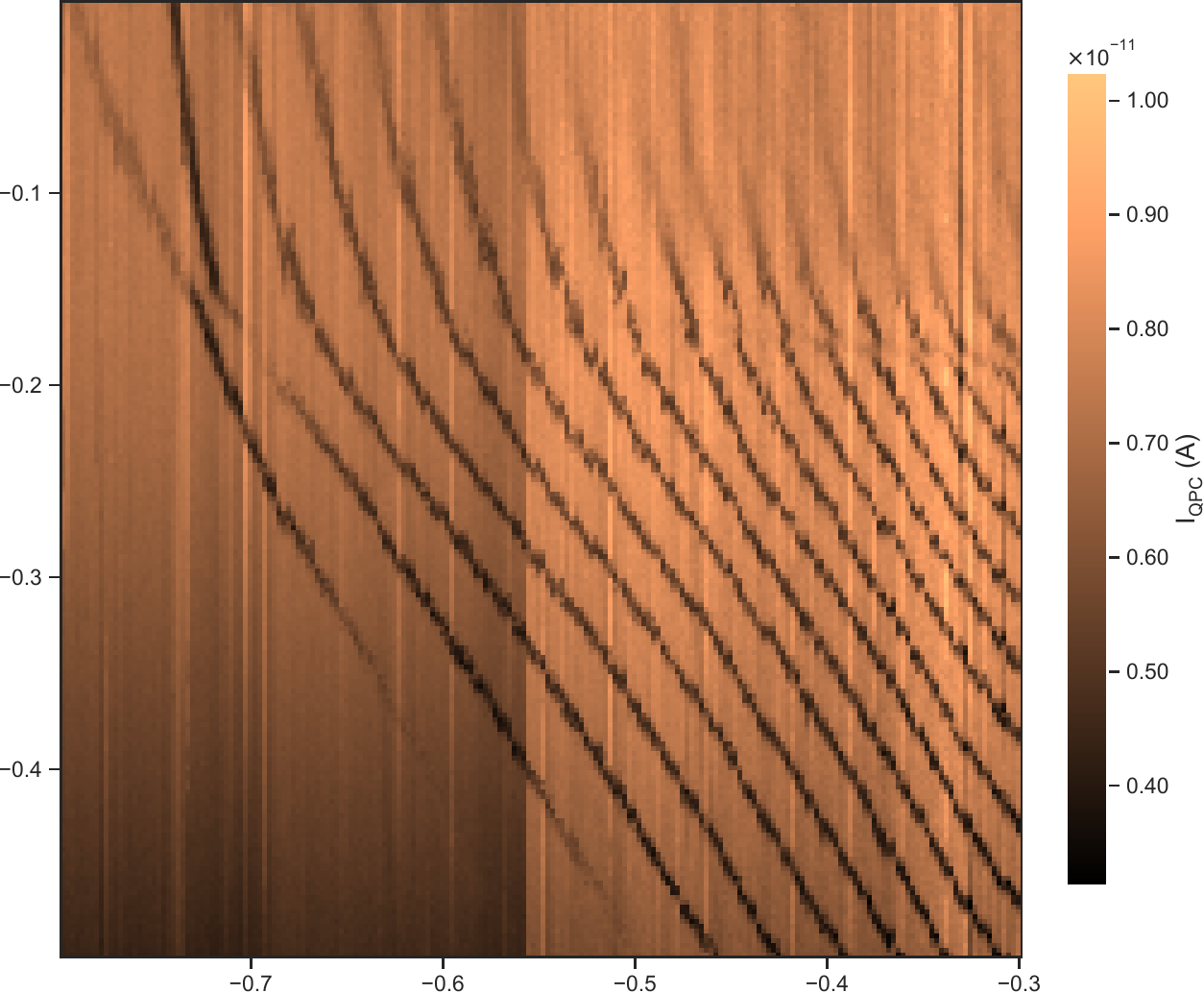}
        \end{subfigure}
        \begin{subfigure}{0.45\textwidth}
            \centering
            \includegraphics[width=\textwidth]{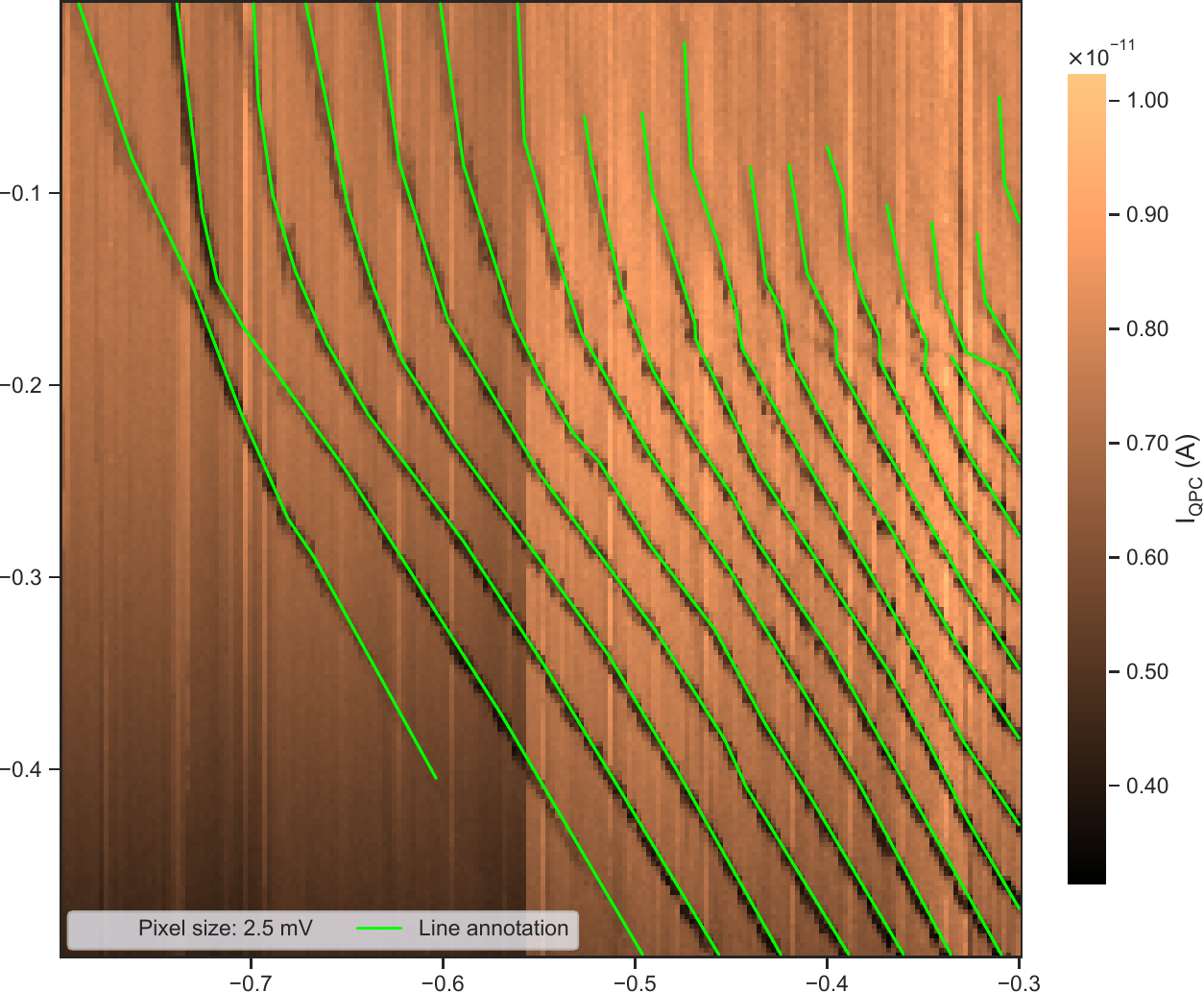}
        \end{subfigure}
        \caption[\acs*{GaAs-QD} stability diagram samples]{
            Two \acf*{GaAs-QD} stability diagram examples, without and with transition line annotations.
            The $x$-axes correspond to the G1 gate voltage, and the $y$-axes correspond to the G2 gate voltage.
        }
        \label{fig:digram-samples-louis}
    \end{figure}

\begin{figure}[H]
        \centering
        \begin{subfigure}{0.45\textwidth}
            \centering
            \includegraphics[width=\textwidth]{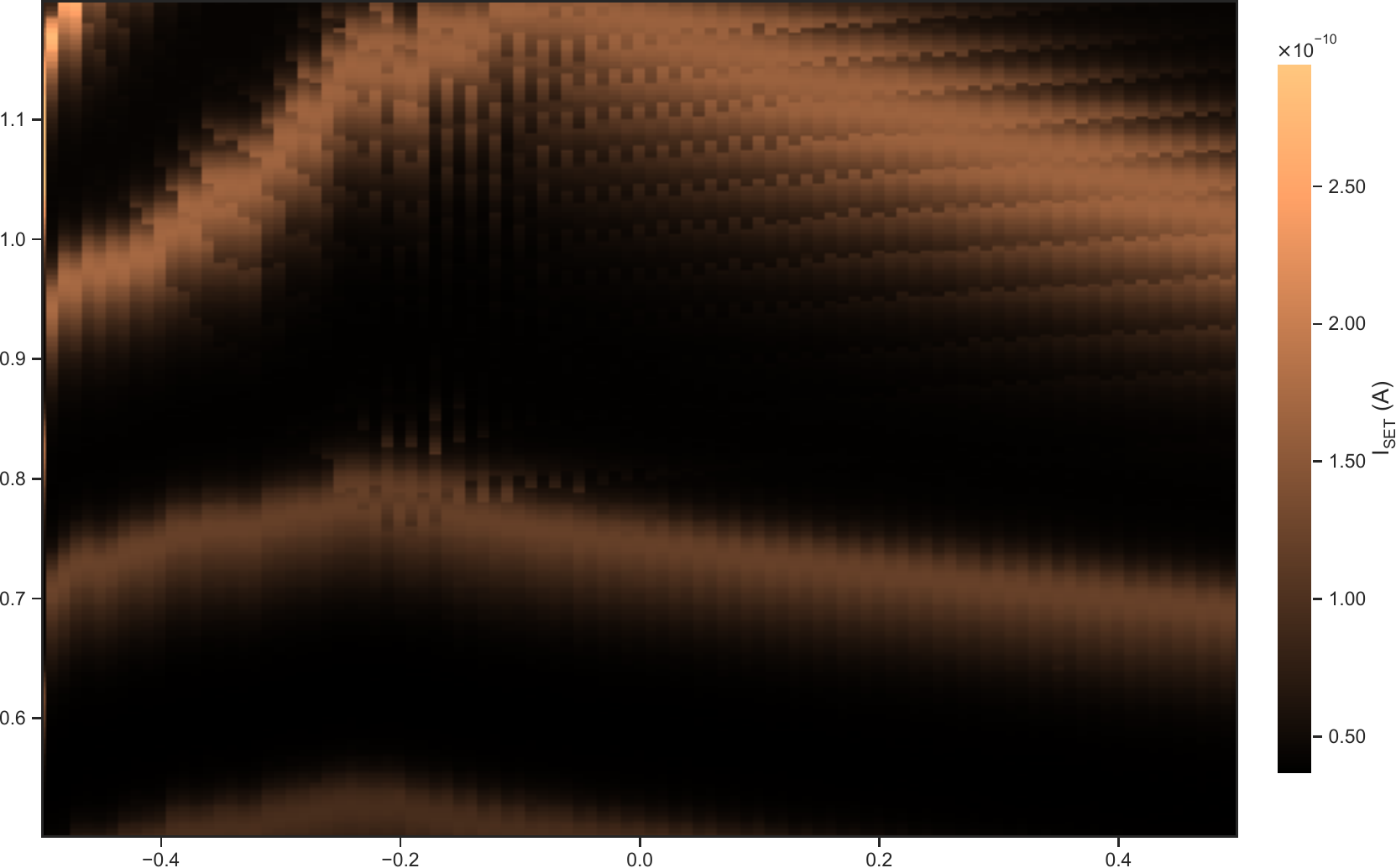}
        \end{subfigure}
        \begin{subfigure}{0.45\textwidth}
            \centering
            \includegraphics[width=\textwidth]{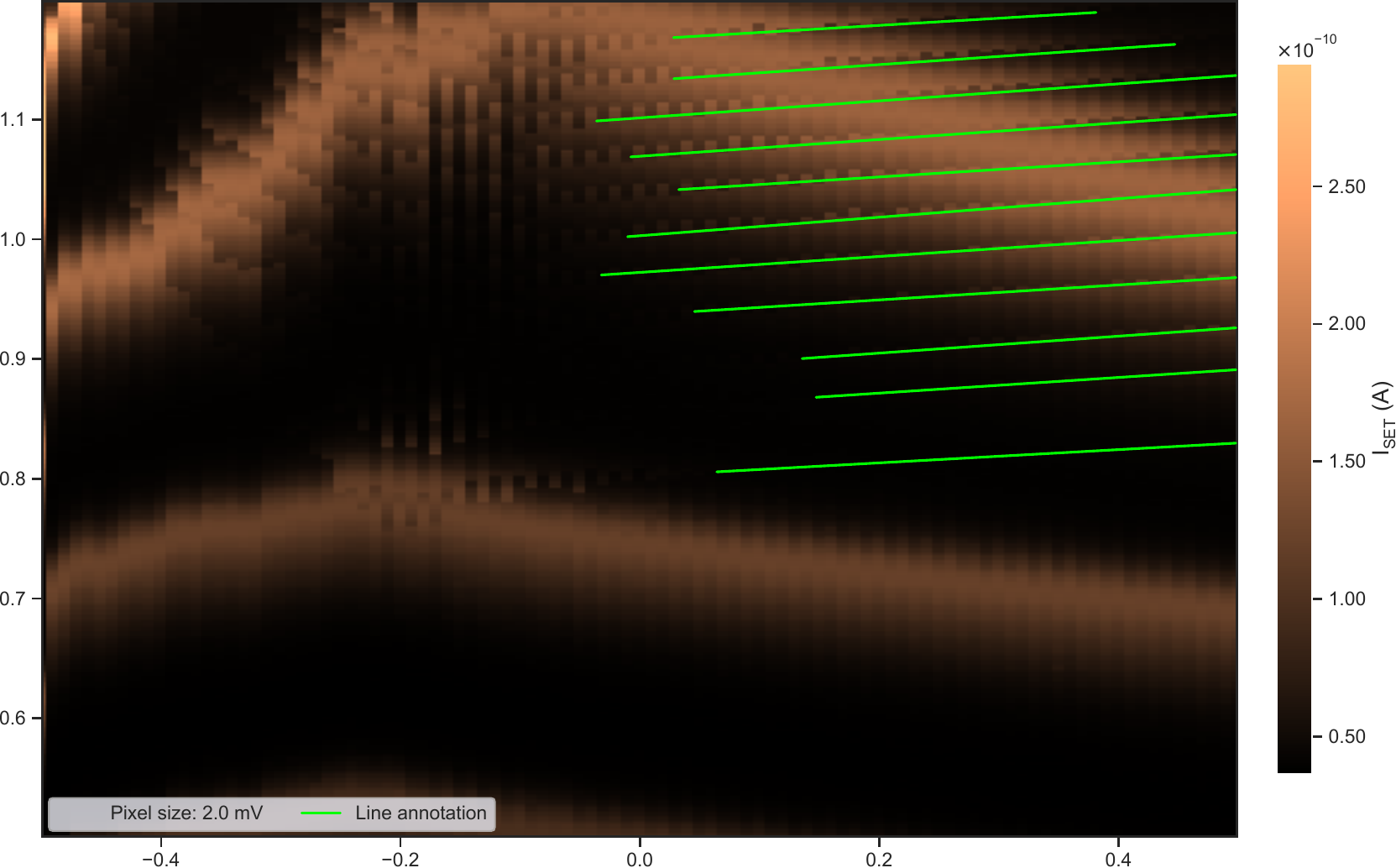}
        \end{subfigure}
        \begin{subfigure}{0.45\textwidth}
            \centering
            \includegraphics[width=\textwidth]{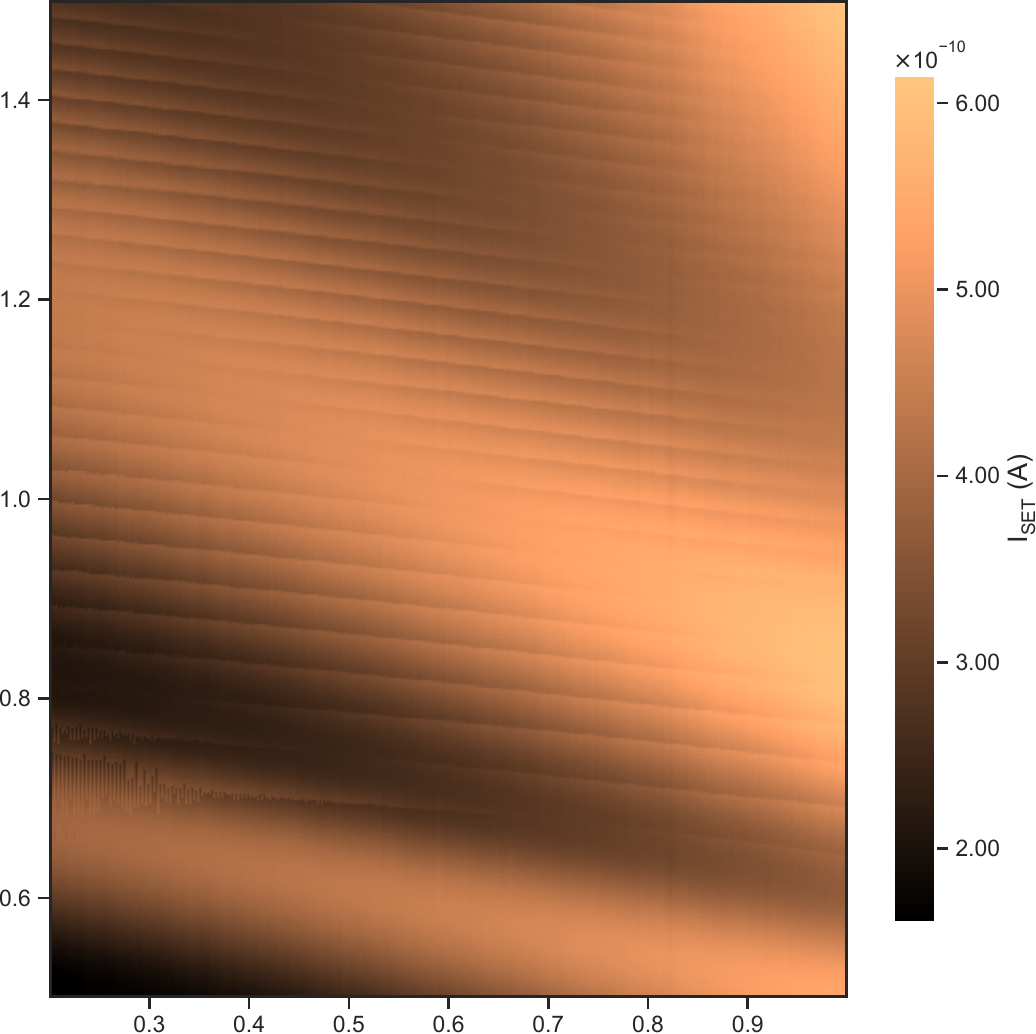}
        \end{subfigure}
        \begin{subfigure}{0.45\textwidth}
            \centering
            \includegraphics[width=\textwidth]{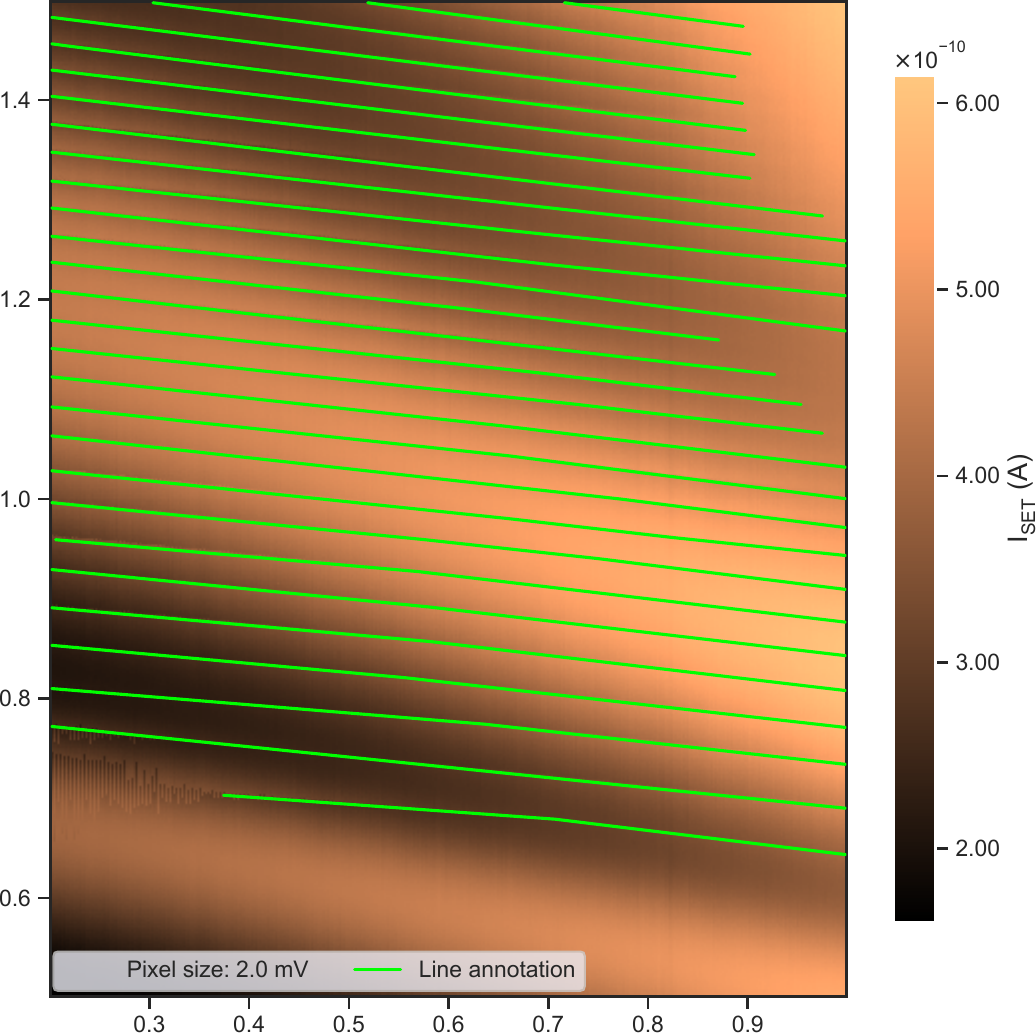}
        \end{subfigure}
        \caption[\acs*{Si-OG-QD} stability diagram samples]{
            Two \acf*{Si-OG-QD} stability diagram examples, without and with transition line annotations.
            The $x$-axes correspond to the G1 gate voltage, and the $y$-axes correspond to the G2 gate voltage.
        }
        \label{fig:digram-samples-eva}
    \end{figure}

    All of stability diagrams used in this study can be downloaded from \citet{qdsd}.
    To illustrate the transition line annotation methodology and demonstrate the diagram diversity of each dataset, two diagrams of each dataset are presented in Figure~\ref{fig:digram-samples-michel}, Figure~\ref{fig:digram-samples-louis}, and Figure~\ref{fig:digram-samples-eva}.

    \section{Line detection methodology}\label{sec:suppl-line-method}

        \begin{table}[H]
        \centering
        \caption[Model architectures and training meta-parameters]{
        	Model architectures and training meta-parameters used in this study.
        }
        \label{tab:model-parameter}
        \begin{tabular}{|c|c|c|c|}
            \hline 
            \textbf{Meta-parameters} & \textbf{\acs{FF}} & \textbf{\acs{CNN}} & \textbf{\acs{BCNN}} \\
            \hline 
            Number of training updates   & \num{15000}             & \multicolumn{2}{c|}{\num{30000}}      \\
            \hline 
            Layers description &
                \makecell{Fully connected:\\
                - \num{400}\\
                - \num{100}} &
                \multicolumn{2}{c|}{
                    \makecell{Convolutions:\\
                    - $4\times4$ kernel, \num{12} channels\\
                    - $4\times4$ kernel, \num{24} channels\\
                    Fully connected:\\
                    - \num{200}\\
                    - \num{100}}
                }\\
            \hline 
            Number of free parameters   & \num{170201}            & \num{367267}       & \num{734534}                     \\
            \hline 
            Loss function               & \multicolumn{2}{c|}{Binary cross-entropy}    & \makecell{
                                                                                            Binary cross-entropy + \\
                                                                                            complexity
                                                                                            cost~\cite{Blundell_2015}
                                                                                        }                                 \\
            \hline 
            Optimizer                   & \multicolumn{3}{c|}{Adam~\cite{Kingma_2014}}                                    \\
            \hline 
            Learning rate               & \num{0.0005}                                 & \multicolumn{2}{c|}{\num{0.001}} \\
            \hline 
            Dropout rate                & \multicolumn{2}{c|}{\qty{60}{\percent}}      & \qty{0}{\percent}                \\
            \hline 
            Batch size                  & \multicolumn{3}{c|}{\num{512}}                                                  \\
            \hline 
            \makecell{Number of inferences per\\
                        patch ($N$ in Equation~\ref{eq:conf-bayes})} & \multicolumn{2}{c|}{\num{1}} & \num{10}            \\
            \hline 
        \end{tabular}
    \end{table}

        \begin{figure}[H]
            \centering
            \includegraphics[width=.85\textwidth]{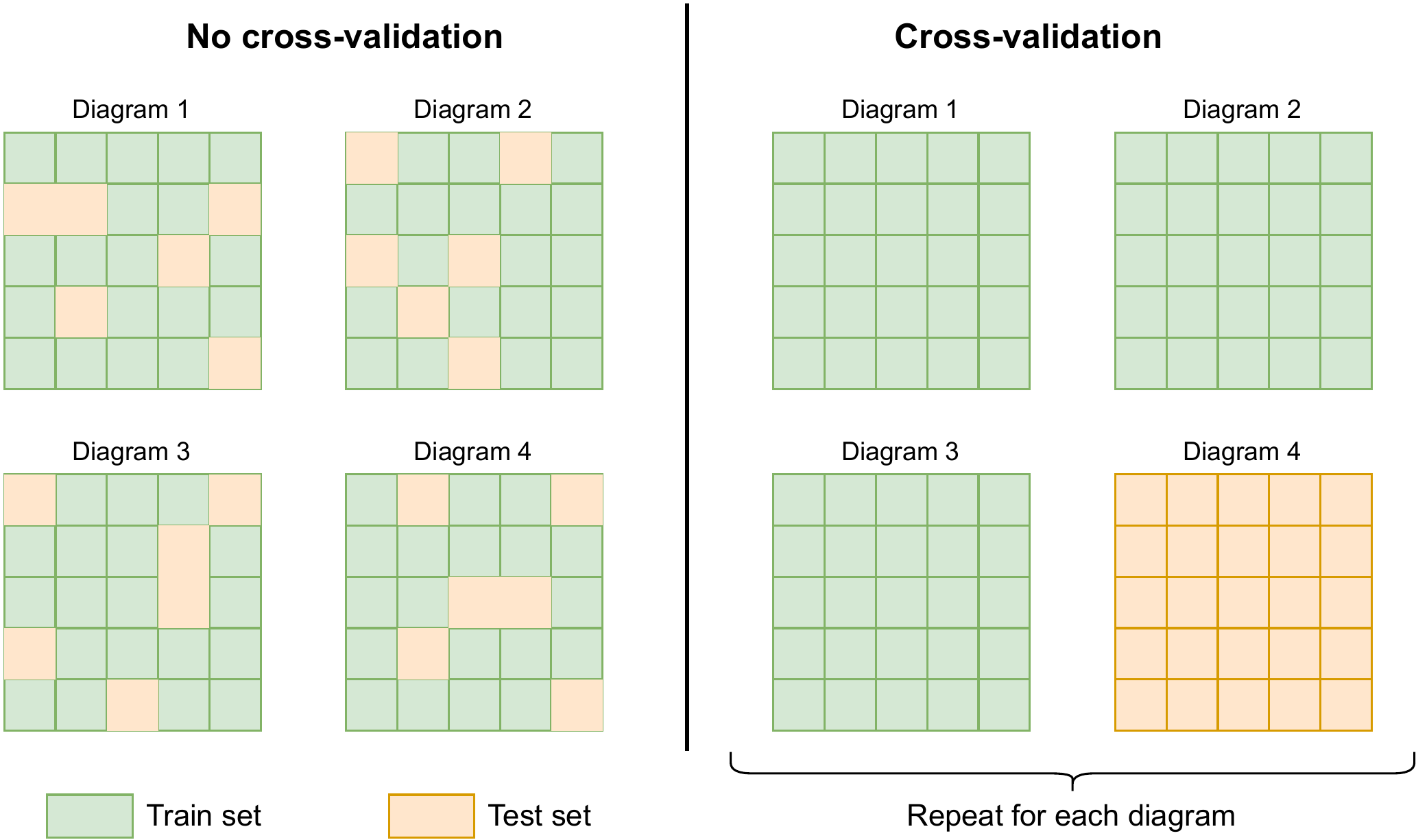}
            \caption[Diagram cross-validation methodology schematic]{
                Diagram cross-validation methodology schematic.
                The left panel represents the ``No cross-validation'' method, where every diagram is split into patches, then a random set of patches is used for training, and the rest are used for testing.
                The right panel represents the ``Cross-validation'' method~\cite{Raschka_2018}, where every diagram is split into patches, then the patches of one diagram are used for testing, while the patches of the others are used for training. This process is repeated for every diagram, and the performance is averaged.
            }
            \label{fig:cross-validation}
        \end{figure}

        \subsection{Model training}\label{subsec:training}

        Each \ac{NN} was trained using PyTorch~\cite{Paszke_2019} v2.1 in a supervised manner.
        We also used the \emph{BLiTZ} library~\cite{Esposito_2020} to create the Bayesian layers, sample the parameters, and compute the parameter updates using \emph{Bayes-by-Backprop}~\cite{Blundell_2015} for the \acp{BCNN}.
        The training parameters and the \ac{NN} architectures are described in Table~\ref{tab:model-parameter}.
        The source code used to obtain the results presented in this article is available on GitHub~\footnote{Git repository: \href{https://github.com/3it-inpaqt/dot-calibration-v2/tree/offline-article}{github.com/3it-inpaqt/dot-calibration-v2/tree/offline-article}}, and the output files (including the trained model parameters) can be downloaded from \citet{run_outputs}.

        The imbalanced class ratio (``\emph{line}'' / ``\emph{no-line}'') caused convergence issues, since the loss could be easily reduced by predicting only the most frequent class (``\emph{no-line}'').
        We worked around this problem by re-balancing the patch dataset using weighted sampling for each batch.

        The meta-parameters presented in Table~\ref{tab:model-parameter} were selected based on a grid search and informed guesses.
        It is likely that these meta-parameters could be optimized and fine-tuned to improve the accuracy, accelerate the training, and reduce the number of free parameters.
        However, the scope of this study is to demonstrate the feasibility of the autotuning procedure and the benefits of using model uncertainty, not to achieve the best possible performance.

        \subsection{Model uncertainty and confidence threshold calibration}\label{subsec:threshold}

        To estimate and use \ac{NN} uncertainty, we made three important decisions: what model to use, what confidence metric to compute, and how to define the confidence threshold under which the inference should not be considered reliable.

        We evaluated a standard \ac{CNN} and \ac{FF} as references, since these are the most simple and well-established \acp{NN} for image pattern recognition.
        Such models are not specifically designed or trained to estimate uncertainty.
        Several studies~\cite{Vasudevan_2019, Hendrycks_2016, Sensoy_2018, Mozejko_2018} have suggested that classical \acp{NN} are generally overconfident and provide badly calibrated confidence scores.
        For the sake of simplicity, we defined the confidence score as the distance between the output and the closest class (Formula~\ref{eq:conf-heuristic}).
        This score is normalized to a range between 0 and 1 to be easily interpretable and comparable as a confidence percentage.

        We chose to implement \iac{BCNN} using variational inference~\cite{Hinton_1993, Graves_2011} and trained it using the Bayes-by-Backprop~\cite{Blundell_2015} method.
        This choice was motivated by previously promising results~\cite{Kwon_2020, Jospin_2022}, and the library~\cite{Esposito_2020} was compatible with PyTorch~\cite{Paszke_2019}.
        The confidence score was computed as a normalized standard deviation (Formula~\ref{eq:conf-bayes}) from \num{10} inferences of the same input with newly sampled parameters.

\begin{figure}[H]
            \centering
            \begin{subfigure}{0.45\textwidth}
                \centering
                \raisebox{12em}{ 
                    \includegraphics[width=\textwidth]{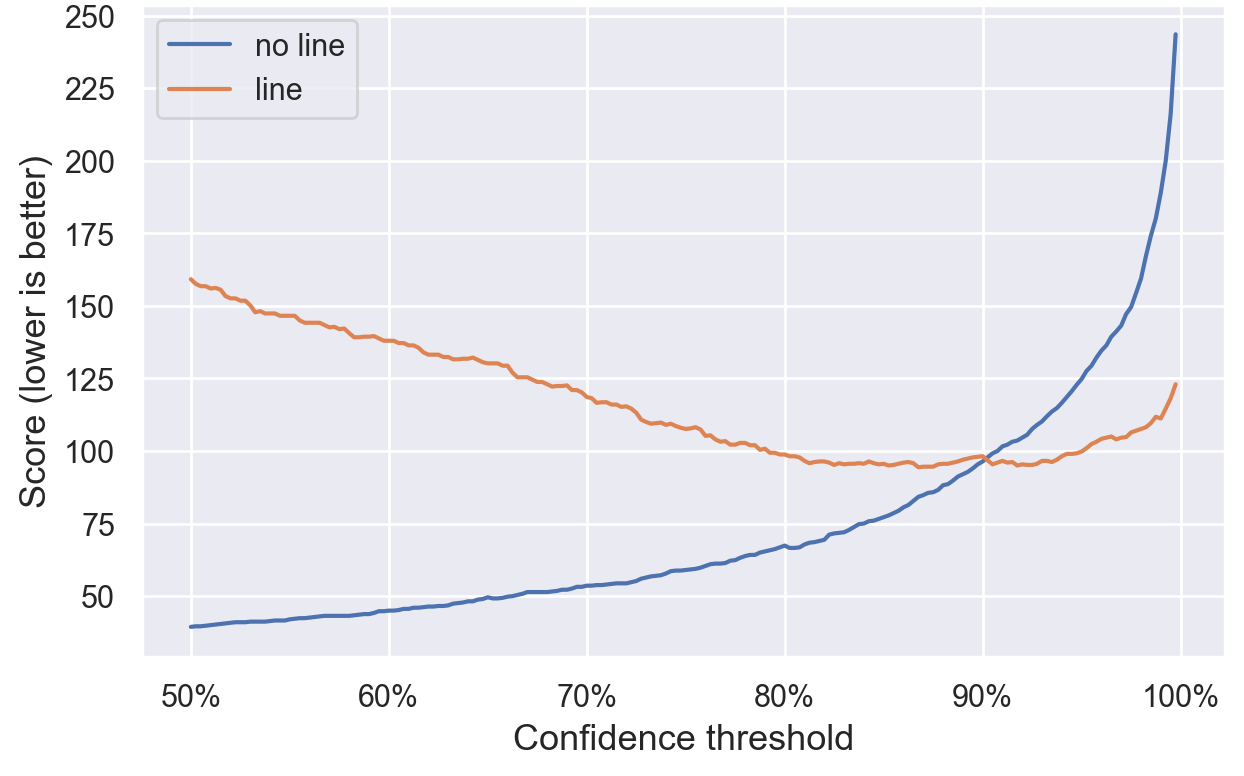}
                }
                \caption{
                    Evolution of the threshold score (Formula~\ref{eq:threshold}) when the threshold is varied.
                    The threshold with the lowest score value is selected as the optimal threshold for each class.
                }
            \end{subfigure}
            \begin{subfigure}{0.50\textwidth}
                \centering
                \includegraphics[width=\textwidth]{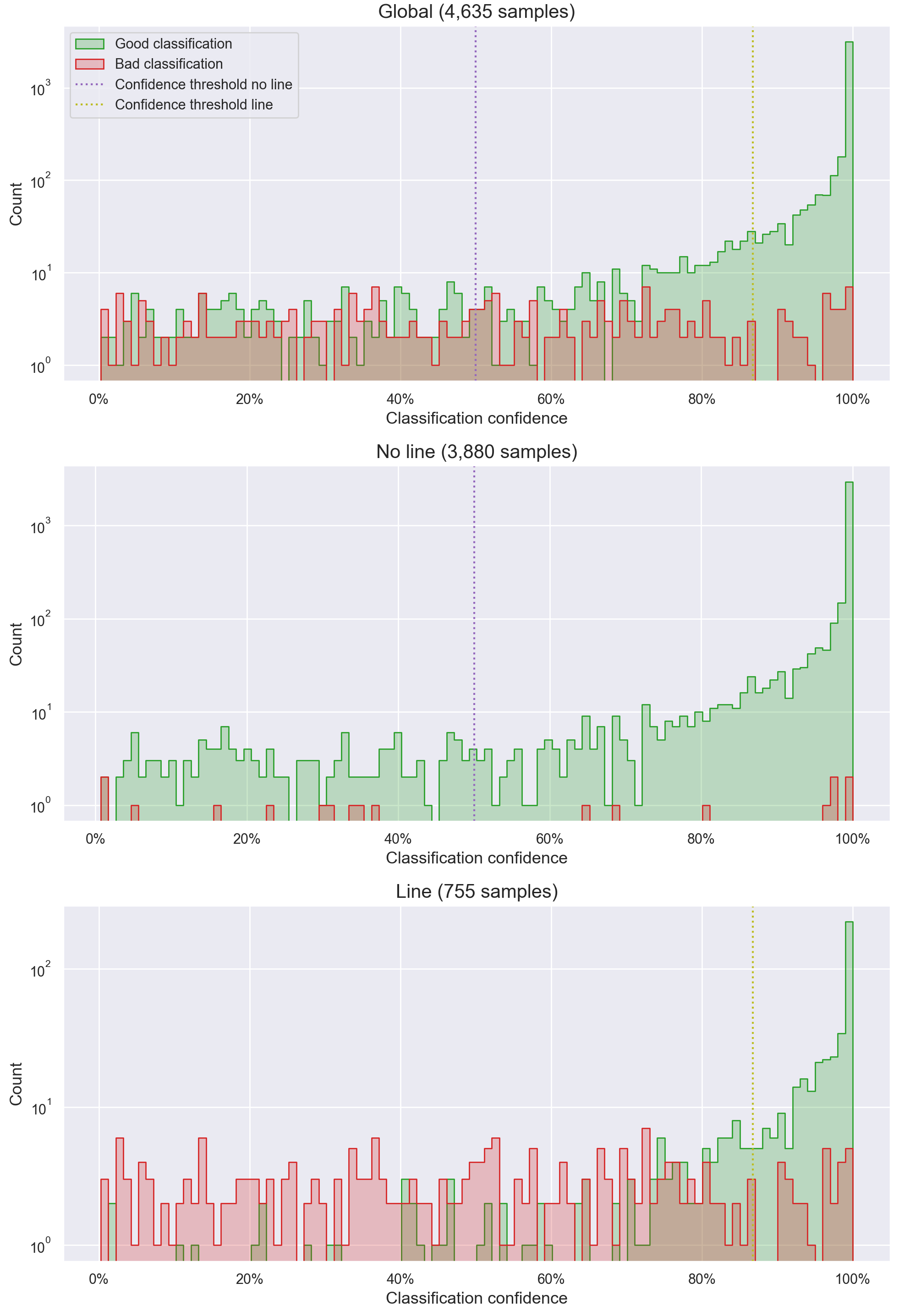}
                \caption{
                    Histogram of the model classification error as a function of the confidence score.
                    The confidence thresholds obtained from (a) are represented by the vertical dotted lines.
                    The top panel shows the distribution for both classes combined, while the bottom panels show the distribution for each class independently.
                }
            \end{subfigure}
            \caption[Confidence thresholds calibration]{
                Example of confidence thresholds calibration for the \acf*{Si-OG-QD} dataset using a \acf*{CNN} and the cross-validation method.
                The thresholds are calibrated after training, on \num{4635} patches from the validation set.
                The confidence threshold of each class is calibrated independently.
                The small number of classification errors for the ``\emph{no-line}'' class in the validation set makes the calibration challenging. Therefore, a value of \qty{50}{\percent} is used as the default minimal threshold.
            }
            \label{fig:threshold-calibration}
        \end{figure}

        A model is well calibrated if the confidence score accurately expresses the probability of making an error; for example, \qty{20}{\percent} of the classifications with \qty{80}{\percent} confidence should be wrong.
        However, a trained \ac{NN} is never perfectly calibrated, and the calibration task is challenging for several reasons.
        First, \ac{NN} training involves some stochastic processes (parameter initialization, stochastic gradient descent, etc.) that lead to different results.
        Therefore, two identical models trained separately could have very different and unpredictable miscalibration issues.
        Secondly, within one model, each predicted class could necessitate independent calibration (Figures~\ref{fig:threshold-calibration}).
        We also suspect that the unbalanced class distribution of the dataset could amplify the problem.
        Finally, calibrating the confidence score using the training set induces a bias, since the out-of-distribution examples are (by definition) not represented.
        Correctly expressing the distributional uncertainty in the model score is critical, especially for the cross-validation testing method.

        Model calibration methods require enough samples to cover the entire confidence spectrum, which might be challenging with a limited dataset.
        For example, a model with a high accuracy will correctly classify most of the samples with a high confidence score, implying that the samples with low confidence intervals will be very sparse (e.g., middle panel of Figure~\ref{fig:threshold-calibration}b).
        Therefore, it is impossible to tell whether the model is under-confident or over-confident in these sparse intervals.
        Fortunately, the presented uncertainty-based exploration strategy does not require the exact probability of correctness for all the confidence ranges.
        The purpose of the confidence score in the context of this exploration task is to optimize the exploration--exploitation tradeoff~\cite{Auer_2002, Simpkins_2008}.
        When classification is performed with a confidence score under the threshold, adjacent patches will be explored.
        If the algorithm explores too much, it may waste time on unnecessary measurements, leading to inefficiency.
        Conversely, if it exploits too soon, it risks making incorrect tuning decisions based on incomplete or inaccurate information.
        In other words, we want to reduce the number of errors (impacting the tuning success rate) relative to the number of patches scanned (impacting the tuning time).
        Calibrating the threshold value by minimizing the score defined by Formula~\ref{eq:threshold} is a simple way to account for the exploration--exploitation tradeoff.
        Fixing the meta-parameter $\tau$ to \num{0.2} is equivalent to saying that we prefer to have up to five more patches scanned rather than having a classification error.

        Other types of models, meta-parameter values, training methods, confidence scores, and calibration methods could be evaluated to address this autotuning problem.
        However, an extensive benchmark of \ac{NN} uncertainty is out of the scope of this paper.

        \subsection{Patch classification samples}\label{subsec:patch-samples}

        \begin{figure}[H]
            \centering
            \begin{subfigure}{0.47\textwidth}
                \centering
                \includegraphics[width=\textwidth]{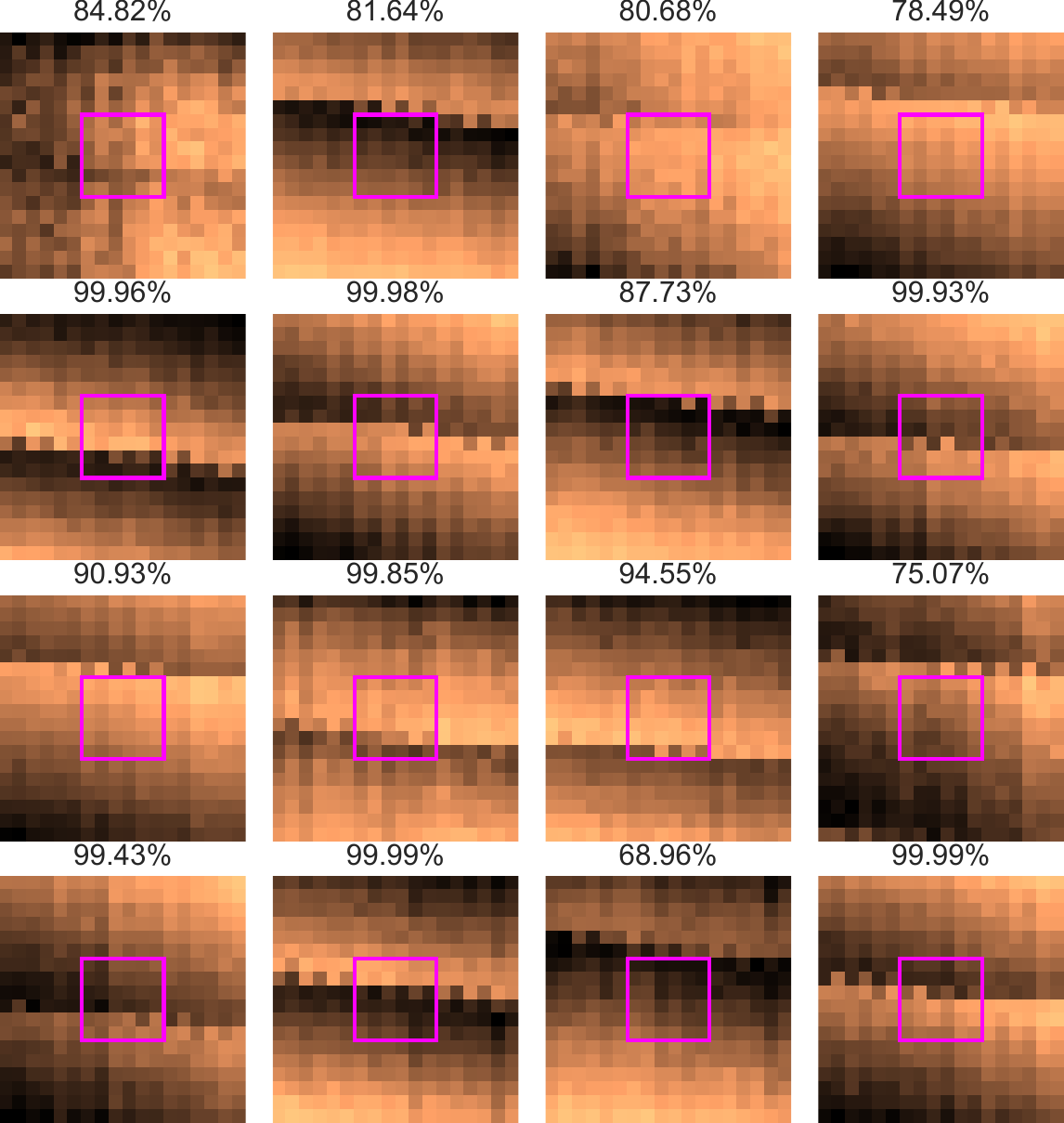}
                \caption{
                    Patches correctly classified as ``\emph{line}'':
                    \num{16} examples among the \num{947} patches in this case.
                }
            \end{subfigure}\hfill
            \begin{subfigure}{0.47\textwidth}
                \centering
                \includegraphics[width=\textwidth]{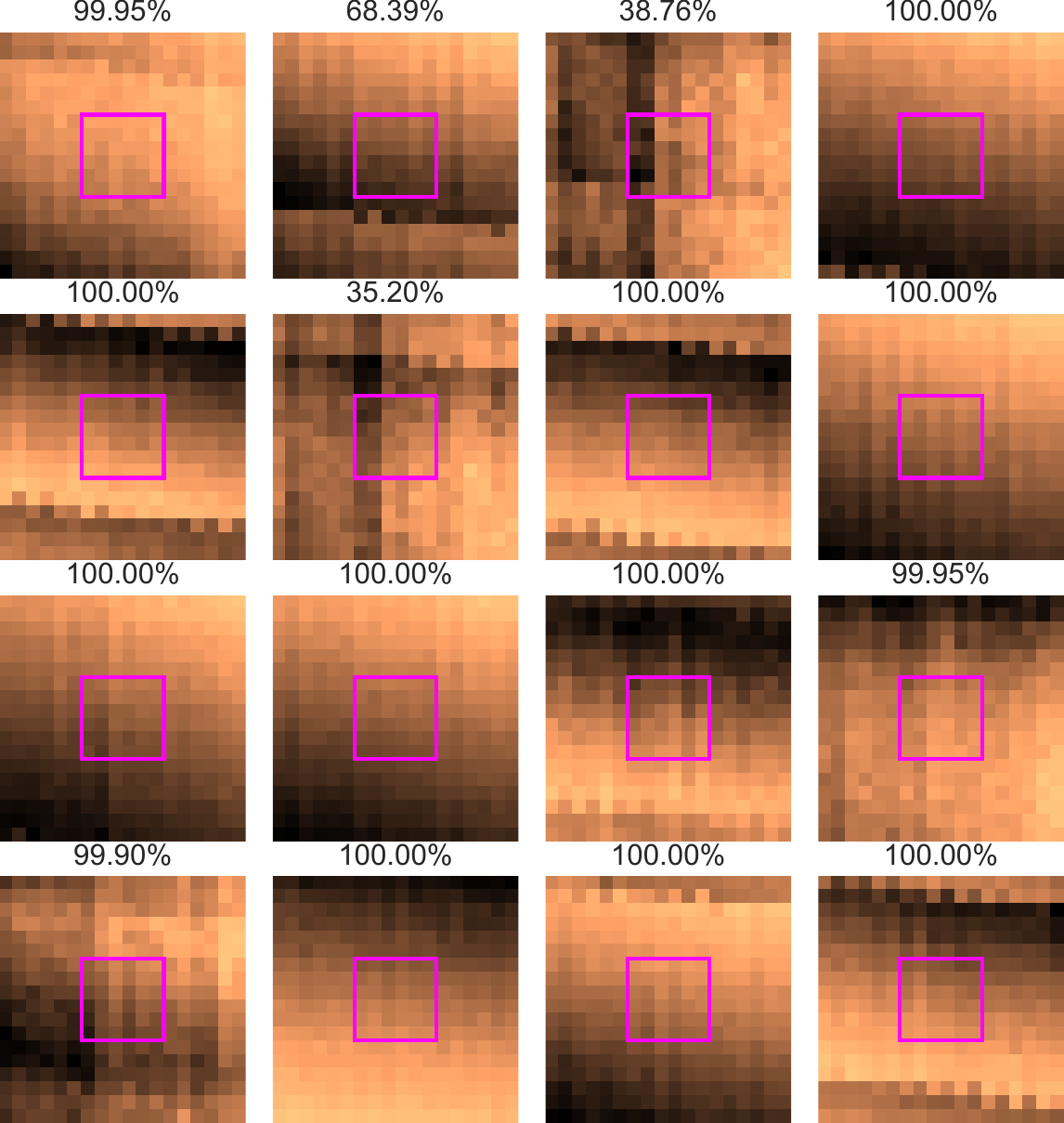}
                \caption{
                    Patches correctly classified as ``\emph{no-line}'':
                    \num{16} examples among the \num{1438} patches in this case.
                }
            \end{subfigure}
            \begin{subfigure}{0.47\textwidth}
                \centering
                \includegraphics[width=\textwidth]{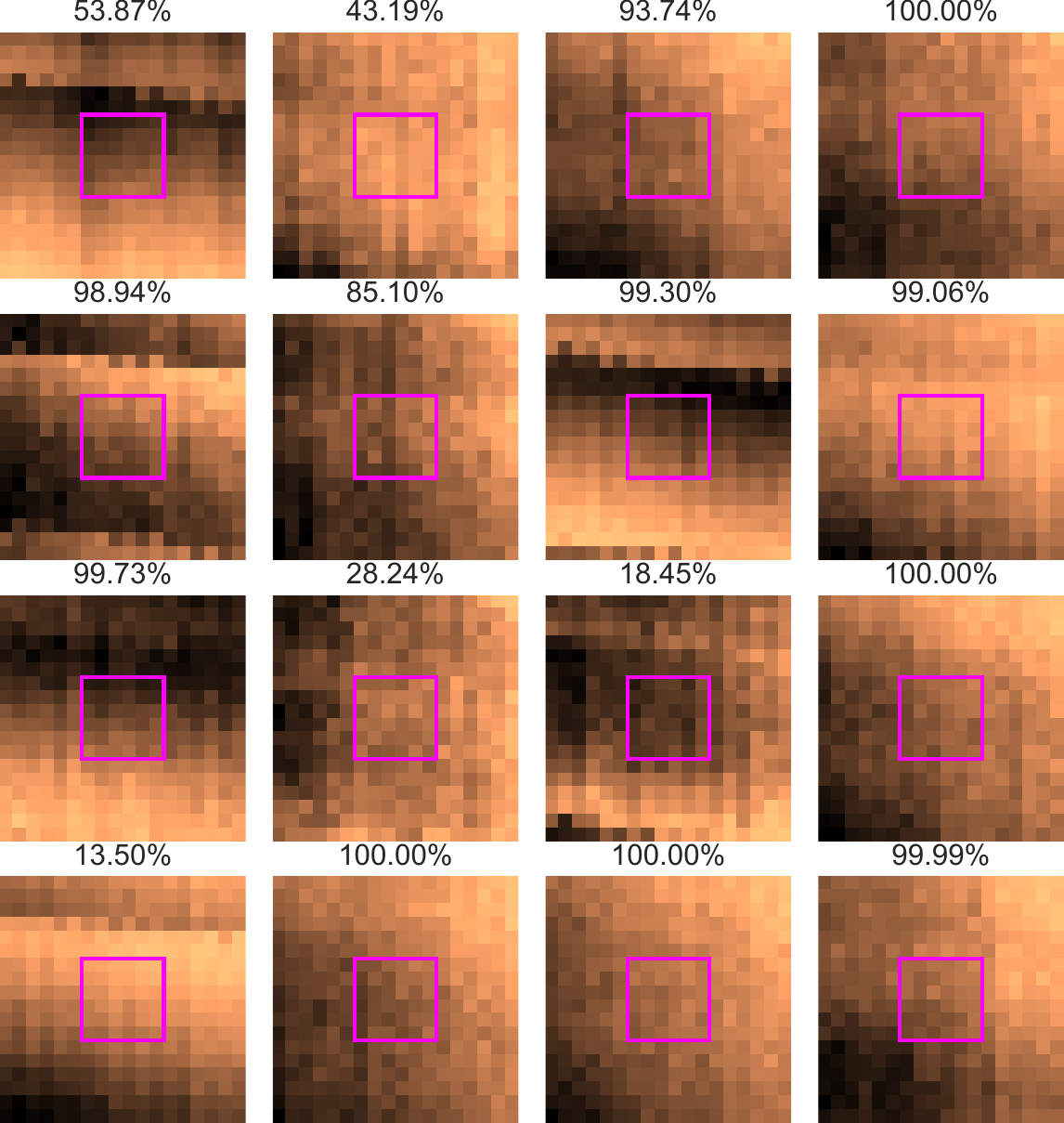}
                \caption{
                    Patches incorrectly classified as ``\emph{no-line}'' (labeled as ``\emph{line}''):
                    \num{16} examples among the \num{75} patches in this case.
                }
            \end{subfigure}\hfill
            \begin{subfigure}{0.47\textwidth}
                \centering
                \includegraphics[width=\textwidth]{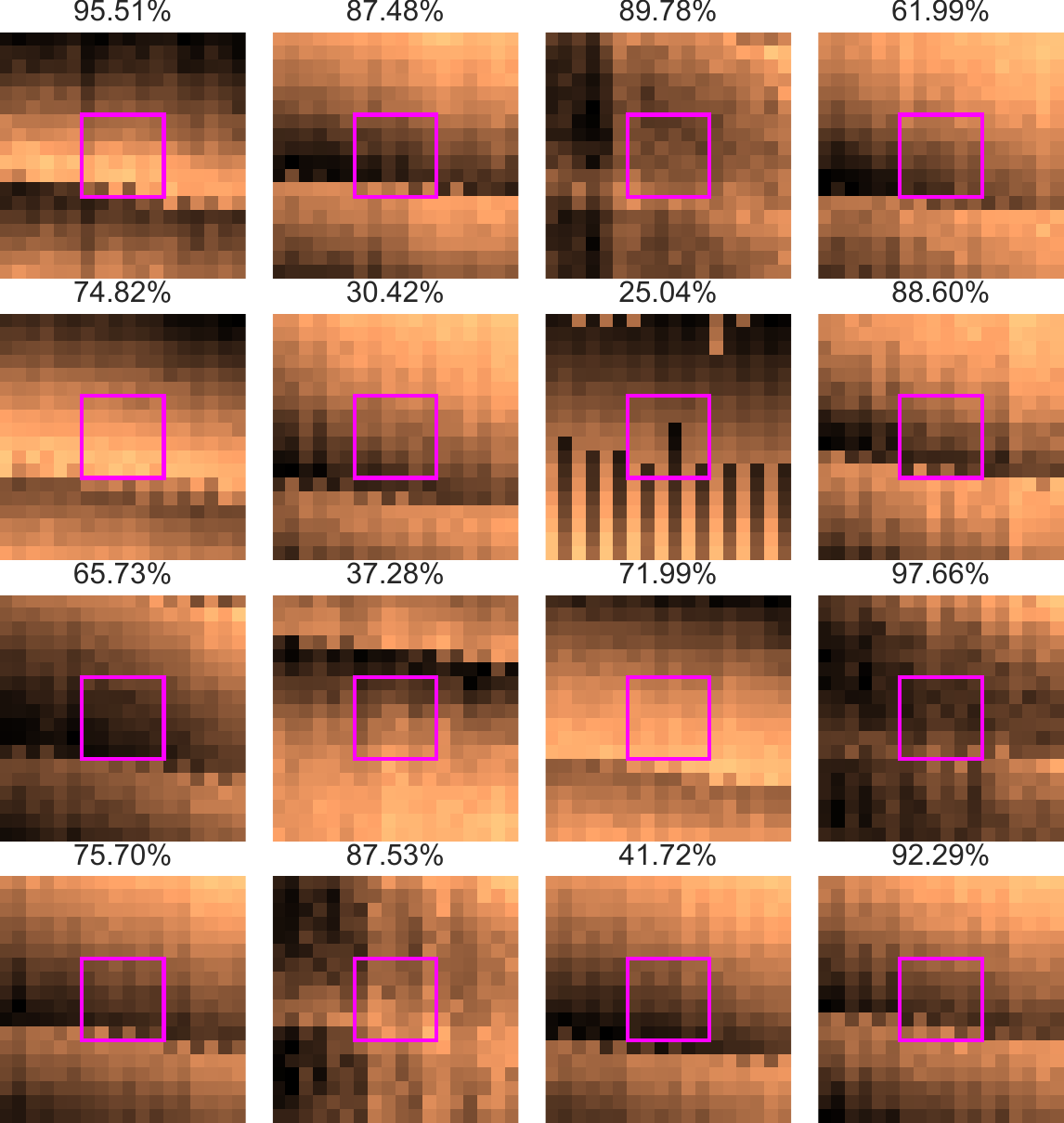}
                \caption{
                    Patches incorrectly classified as ``\emph{line}'' (labeled as ``\emph{no-line}''):
                    \num{16} examples among the \num{468} patches in this case.
                }
            \end{subfigure}
            \caption[Patch samples]{
                Examples of patch classifications for the \acf*{Si-OG-QD} dataset using a \acf*{CNN} and the cross-validation training method.
                The patches are randomly selected among the test set.
                The pink squares in the center of the patches correspond to the detection area.
                A patch is labeled as ``\emph{line}'' only if a transition line annotation crosses this square.
                The percentages at the top of each patch correspond to the confidence scores of the \acs*{CNN} computed using Formula~\ref{eq:conf-heuristic}.
            }
            \label{fig:patch-samples}
        \end{figure}

        Figure~\ref{fig:patch-samples} presents examples of the classification results to illustrate the line detection performance.
        We can observe that the classification errors (Figures~\ref{fig:patch-samples}c,d) are often due to the presence of a transition line close to the detection area or annotation mistakes.
        In general, the confidence scores are lower when the patch is misclassified.
        More samples can be found in the simulation result files available for download from \citet{run_outputs}.

    \section{Autotuning methodology}\label{sec:suppl-autotuning-method}

        \subsection{Exploration algorithm}\label{subsec:suppl-exploration-algo}

\begin{figure}[H]
            \centering
            \includegraphics[width=.95\textwidth]{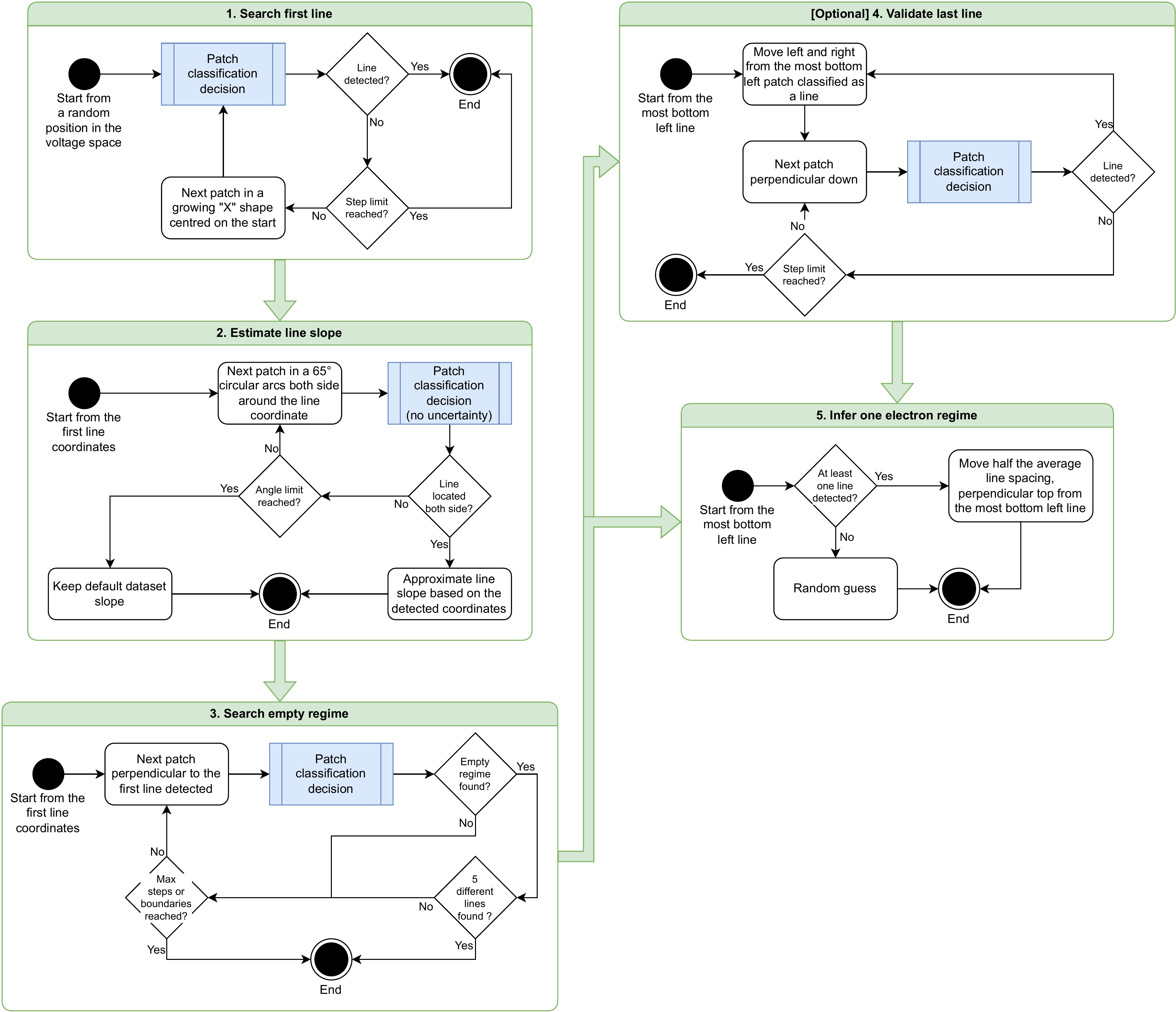}
            \caption[Detailed exploration strategy schematic]{
                Detailed exploration strategy schematic representing each step of the autotuning procedure.
                The optional \emph{step 4} is enabled for the \acs*{GaAs-QD} and \acs*{Si-OG-QD} datasets to improve the tuning success rate when the first transition line is fading (see examples in Figure~\ref{fig:digram-samples-eva}).
            }
            \label{fig:tuning-algo-suppl}
        \end{figure}

        The autotuning exploration algorithm is illustrated as a block diagram in Figure~\ref{fig:tuning-algo-suppl}.

        \subsubsection*{Notes about the estimation of the transition line slope (step 2)}

        After finding the first patch containing a transition line, we scanned and classified a sequence of patches in circular arcs (up to \ang{65}) on both sides surrounding this location.
        This scan sequence is illustrated in step 2 of Figure~\ref{fig:tuning-algo} (the white arrows emphasize the arc directions).

        We consider an arc to be valid if we detect a sequence of patches matching the following pattern:

        \[m \times \textit{no-line}, n \times \textit{line}, p \times \textit{no-line} \]

        where $m$, $n$, and $p$ are positive integers representing the number of consecutive patches classified in the given label.

        If both arcs are valid, we can estimate the coordinates of two points in the line by taking the position between the first and last patch coordinates in the $n$-th sequence of each arc.
        Using these two points and the \emph{2-argument arctangent} function, we can extract the estimated slope value.

        If at least one of the arcs is invalid (e.g., the line fades or a patch is misclassified), we keep the default slope value until the end of the run.

        \subsection{Baselines}\label{subsec:suppl-baselines}

        Two simple baselines are used as references to interpret and analyze the results of the autotuning simulations.
        The benchmark methodology is identical to the other autotuning procedures, including the repetition over \num{10} random seeds and \num{50} random starting points for each stability diagram.

        The ``\emph{oracle}'' baseline uses the exact same exploration logic as the one presented in this article, but the line detection model is replaced by an ``\emph{oracle}'' that has access to the ground truth labels.
        Therefore, every patch classification is \qty{100}{\percent} accurate, with a maximal confidence score.
        This baseline allows us to isolate the tuning errors related to the exploration strategy from the ones caused by model misclassification.
        The results of this baseline corroborate the conclusion that the \ac{GaAs-QD} and \ac{Si-OG-QD} datasets are more challenging to explore, possibly due to the transition line irregularities and unpredictable shapes.

        The ``\emph{random}'' baseline represents the worst case of exploration, where a pair of coordinates is randomly selected inside the voltage space of the tuned stability diagram.
        The success rate of this baseline can be interpreted as the average surface area of the one-electron regime in the dataset's diagrams.
        A larger surface is expected to be easier to find using an autotuning procedure.

        Therefore, these two baselines defined the minimal and maximal tuning success rate and average number of steps expected for each dataset.

        \begin{table}[H]
            \centering
              \caption{
            	Autotuning results using ``\emph{oracle}'' and ``\emph{random}' baselines.
            	Variability over \num{10} random seeds.
            }
            \label{tab:baselines}
            \begin{tabular}{|c|c|cc|}
                \hline 
                Dataset &
                Baseline &
                \makecell{Average step\\number} &
                \makecell{Tuning success} \\
                \hline 
                                          & Oracle & 151 & 100.0\% \std{0.1} \\
                \multirow{-2}{*}{\michel} & Random & 0   &  11.8\% \std{1.5} \\
                \hline 
                                          & Oracle & 94  &  96.7\% \std{0.6} \\
                \multirow{-2}{*}{\louis}  & Random & 0   &  10.1\% \std{1.4} \\
                \hline 
                                          & Oracle & 164 &  92.0\% \std{0.9} \\
                \multirow{-2}{*}{\eva}    & Random & 0   &   5.3\% \std{1.0} \\
                \hline 
            \end{tabular}
          \end{table}

    \section{Results without cross-validation}\label{sec:suppl-no-cross-validation}

    The performance loss observed when using the cross-validation method (illustrated in Figure~\ref{fig:cross-validation}) gives us insight into the effects of the distributional uncertainty on the autotuning success rate.
    The results of the line detection model trained without cross-validation, presented in Table~\ref{tab:line-results-no-cross-validation}, show high accuracy for every dataset (above \qty{98}{\percent} when using the uncertainty threshold) and a consistently lower rate of instances below the threshold (number of patches classified as ``\emph{unknown}'') compared to the model trained with cross-validation (presented in Table~\ref{tab:line-results}).
    As shown in Table~\ref{tab:results-no-cross-validation}, the autotuning success rates directly benefit from the model performance gain.
    This gap is even greater when the dataset contains very different diagrams, such as in the \ac{GaAs-QD} dataset.
    Therefore, reducing the distributional uncertainty seems to be a promising way to improve the autotuning success rate, which could be achieved either by covering a boarder range of diagrams in the training set or by reducing the device fabrication variability.

    \begin{table}[H]
        \centering
        \caption[Transition line detection results without cross-validation]{
        	Line detection results without cross-validation for each dataset and model.
        	The performances are averaged over \num{10} runs with different random seeds.
        	The standard deviation of the run performances represents the variability of the methods.
        	Each run covers every diagram of the dataset.
        	The best test accuracy scores of each dataset are highlighted in bold.
        }
        \label{tab:line-results-no-cross-validation}
        \begin{tabular}{|c|c|cc|cc|}
            \hline 
            Dataset &
            Model &
            Accuracy &
            \makecell{Accuracy above\\threshold} &
            \makecell{Error reduction\\using threshold} &
            \makecell{Rate below\\threshold} \\
            \hline 
                                      & BCNN & 98.8\% \std{0.1} & \textbf{99.8\%} \std{0.1} & 79.9\% \std{5.1}  & 2.0\% \std{0.3}  \\
                                      & CNN  & 98.7\% \std{0.1} & \textbf{99.8\%} \std{0.1} & 86.4\% \std{3.6}  & 2.0\% \std{0.2}  \\
            \multirow{-3}{*}{\michel} & FF   & 96.5\% \std{0.7} & \textbf{99.8\%} \std{0.1} & 92.9\% \std{3.1}  & 10.7\% \std{2.3} \\
            \hline 
                                      & BCNN & 94.3\% \std{0.9} & 96.9\%          \std{1.1} & 46.5\% \std{15.9} & 8.0\% \std{3.1}  \\
                                      & CNN  & 95.2\% \std{0.7} & \textbf{98.0\%} \std{0.9} & 59.7\% \std{14.3} & 7.5\% \std{2.6}  \\
            \multirow{-3}{*}{\louis}  & FF   & 93.9\% \std{1.1} & 97.6\%          \std{0.8} & 60.9\% \std{10.0} & 10.5\% \std{2.4} \\
            \hline 
                                      & BCNN & 95.3\% \std{0.3} & 98.4\%          \std{0.2} & 65.8\% \std{5.8}  & 7.4\% \std{1.4}  \\
                                      & CNN  & 93.6\% \std{0.4} & \textbf{99.1\%} \std{0.2} & 85.2\% \std{2.3}  & 11.4\% \std{0.9} \\
            \multirow{-3}{*}{\eva}    & FF   & 94.1\% \std{0.3} & 98.9\%          \std{0.2} & 81.0\% \std{3.0}  & 11.1\% \std{0.7} \\
            \hline 
        \end{tabular}
    \end{table}

    \begin{table}[H]
        \centering
         \caption{
        	Autotuning results for each dataset and model without cross-validation, with and without using the model uncertainty information provided by the confidence score.
        	The line detection accuracy and tuning success variability are computed over \num{10} runs with different random seeds.
        	Each run covers every diagram of the dataset.
        	The best tuning success rates of each dataset are highlighted in bold.
        }
        \label{tab:results-no-cross-validation}
        \rowcolors{2}{}{lightgray!30}  
        \begin{tabular}{|>{\cellcolor{white}}c|>{\cellcolor{white}}c|>{\cellcolor{white}}c|ccc|}
            \hline 
            Dataset &
            Model &
            \makecell{Line detection\\accuracy} &
            \makecell{Uncertainty-\\based tuning} &
            \makecell{Average step\\number} &
            Tuning success \\
            \hline 
                                      &                        &                                    & Yes & 166 & 99.5\%          \std{ 0.3} \\
                                      & \multirow{-2}{*}{BCNN} & \multirow{-2}{*}{98.8\% \std{0.1}} & No  & 151 & 92.2\%          \std{ 2.7} \\
                                      &                        &                                    & Yes & 166 & \textbf{99.9\%} \std{ 0.2} \\
                                      & \multirow{-2}{*}{CNN}  & \multirow{-2}{*}{98.7\% \std{0.1}} & No  & 152 & 91.2\%          \std{ 3.5} \\
                                      &                        &                                    & Yes & 199 & 71.2\%          \std{11.7} \\
            \multirow{-6}{*}{\michel} & \multirow{-2}{*}{FF}   & \multirow{-2}{*}{96.5\% \std{0.7}} & No  & 128 & 30.7\%          \std{15.0} \\
            \hline 
                                      &                        &                                    & Yes & 101 & 85.4\%          \std{ 9.9} \\
                                      & \multirow{-2}{*}{BCNN} & \multirow{-2}{*}{94.3\% \std{0.9}} & No  &  93 & 66.2\%          \std{ 6.8} \\
                                      &                        &                                    & Yes & 100 & \textbf{92.2\%} \std{ 2.5} \\
                                      & \multirow{-2}{*}{CNN}  & \multirow{-2}{*}{95.2\% \std{0.7}} & No  &  95 & 85.8\%          \std{ 5.5} \\
                                      &                        &                                    & Yes & 104 & 78.0\%          \std{ 5.6} \\
            \multirow{-6}{*}{\louis}  & \multirow{-2}{*}{FF}   & \multirow{-2}{*}{93.9\% \std{1.1}} & No  &  92 & 63.0\%          \std{ 3.8} \\
            \hline 
                                      &                        &                                    & Yes & 184 & 81.8\%          \std{ 2.3} \\
                                      & \multirow{-2}{*}{BCNN} & \multirow{-2}{*}{95.3\% \std{0.3}} & No  & 160 & 70.2\%          \std{ 3.5} \\
                                      &                        &                                    & Yes & 189 & 84.3\%          \std{ 3.0} \\
                                      & \multirow{-2}{*}{CNN}  & \multirow{-2}{*}{93.6\% \std{0.4}} & No  & 153 & 61.7\%          \std{ 4.3} \\
                                      &                        &                                    & Yes & 194 & \textbf{85.0\%} \std{ 1.1} \\
            \multirow{-6}{*}{\eva}    & \multirow{-2}{*}{FF}   & \multirow{-2}{*}{94.1\% \std{0.3}} & No  & 154 & 65.5\%          \std{ 2.4} \\
            \hline 
        \end{tabular}
    \end{table}

\end{document}